\documentclass[sort&compress,5p,preprint]{elsarticle} 

\usepackage{bp-listings}
\usepackage{hyperref}
\usepackage{url}

\usepackage{calc}
\usepackage{pgf-playgo-lsc}
\usepackage{pgf-bp}

\usepackage{booktabs}       
\usepackage{amsmath,mathtools,amssymb}
\usepackage{amsthm}
\newtheoremstyle{angularbreak}
  {\topsep}
  {\topsep}
  {\normalfont\itshape}
  {0pt}
  {\bfseries}
  {.}
  { }
  {\thmname{#1}~\thmnumber{#2}\thmnote{ (#3)}}
\theoremstyle{angularbreak} 
\usepackage{amsfonts}       
\usepackage{nicefrac}       
\usepackage{microtype}      
\usepackage{adjustbox}
\usepackage{multirow}
\usepackage{subfig}
\usepackage{enumitem}
\usepackage{tikz}
\usetikzlibrary{arrows,automata,positioning,shapes,patterns.meta}

\tikzset{my lts style/.style={
    initial text=$ $,
    every state/.style={
        rectangle,
        align=left,
        text width=19mm,
        rounded corners,
    },
    node distance=4cm,
    ->,
    >=open triangle 45,
    auto,
    transform shape
    }
}

\tikzset{my cbt style/.style={
every state/.style={
    rectangle,
    align=left,
    text width=25mm,
    rounded corners,
},
node distance=3cm,
->,
>=open triangle 45,
auto,
transform shape
}}
\newcounter{magicrownumbers}%
\newcommand\rownumber{\stepcounter{magicrownumbers}\arabic{magicrownumbers}}%
\usepackage{tabularx}
\newcolumntype{R}[1]{>{\hsize=#1\hsize\noindent\raggedleft\arraybackslash}X}%
\newcolumntype{C}[1]{>{\hsize=#1\hsize\centering\arraybackslash}X}%
\newcolumntype{L}[1]{>{\hsize=#1\hsize\noindent\raggedright\arraybackslash}X}%

\newtheorem{definition}{Definition}

\newcommand{\scaletoline}[1]{\maxsizebox{\linewidth}{!}{#1}}

\newcommand{\golcbt}[4][my cbt style,
every state/.style={
    rectangle,
    align=left,
    text width=27mm,
    rounded corners,
},
node distance=4.8cm,
]{\cbt[#1]{#2}{$Query=Q_{#2}$}{#4}}

\newcommand{\cbt}[4][my cbt style]{
#3

$BT_{#2}=$\\
\scaletoline{
\noindent\begin{tikzpicture}[#1]
    \node(desc) {};
    #4
\end{tikzpicture}
}}

\def\tickPartedQuery[#1]{\partedQuery[#1]{tick=1}}

\def\partedQuery[#1]#2{
\left\{\begin{array}{@{}l@{}l}
    #1  &,  #2 \\ 
    \emptyset &, \text{otherwise}
\end{array}\right.}

\journal{Information and Software Technology}

\begin{document}
\begin{frontmatter}
\title{Context-Oriented Behavioral Programming\tnoteref{t1}}
\tnotetext[t1]{This work was partially supported by the Institute for Innovation in Transportation, Tel-Aviv University, and the Fuel-Choices and Smart-Mobility Initiative, Israel.}
\author{Achiya Elyasaf}
\ead{achiya@bgu.ac.il}
\address{Department of Software and Information Systems Engineering,\\Ben-Gurion University of the Negev, Israel}



\begin{abstract}
\textbf{Context:} Modern systems require programmers to develop code that dynamically adapts to different contexts, leading to the evolution of new \emph{context-oriented} programming languages. These languages introduce new software-engineering challenges, such as: how to maintain the separation of concerns of the codebase? how to model the changing behaviors? how to verify the system behavior? and more.

\noindent\textbf{Objective:} This paper introduces \emph{Context-Oriented Behavioral Programming} (COBP) --- a novel paradigm for developing context-aware systems, centered on natural and incremental specification of context-dependent behaviors. As the name suggests, we combine \emph{behavioral-programming} (BP) --- a scenario-based modeling paradigm --- with context idioms that explicitly specify when scenarios are relevant and what information they need. The core idea is to connect the behavioral model with a data model that represents the context, allowing an intuitive connection between the models via update and select queries. Combining behavioral-programming with context-oriented programming brings the best of the two worlds, solving issues that arise when using each of the approaches in separation.

\noindent\textbf{Method:} We begin with providing abstract semantics for COBP and two implementations for the semantics, laying the foundations for applying reasoning algorithms to context-aware behavioral programs. Next, we exemplify the semantics with formal specifications of systems, including a variant of Conway's \emph{Game of Life}. Then, we provide two case studies of real-life context-aware systems (one in robotics and another in IoT) that were developed using this tool. Throughout the examples and case studies, we provide design patterns and a methodology for coping with the above challenges.

\noindent\textbf{Result:} The case studies show that the proposed approach is applicable for developing real-life systems, and presents measurable advantages over the alternatives --- behavioral programming alone and context-oriented programming alone. 

\noindent\textbf{Conclusion:} We present a paradigm allowing programmers and system engineers to capture complex context-dependent requirements and align their code with such requirements. 
\end{abstract}
\end{frontmatter}

\section{Introduction}
\label{sec:intro}
Requirements of context-aware system are often referring to the system context, either by subjecting a requirement to a specific context (e.g., ``do not use the GPS when battery is low'', etc.), or by specifying how the system should interact with the context (e.g., ``change to emergency mode upon fire detection'', etc.). This work formalizes and generalizes an approach for developing context-aware systems, by combining scenario-based programming with context. Specifically, we propose to add explicit idioms for referencing of context in behavioral programming, as elaborated next.

\emph{Behavioral Programming} (BP)~\cite{Harel2012BehavioralProgramming, Harel2010ProgrammingCoordinated} is a language-independent paradigm for programming reactive systems, designed to allow for a natural and incremental specification of behavior. A behavioral program is comprised of a set of scenarios (that say what to do) and anti-scenarios (that say what not to do), that are interwoven at run-time to generate a combined reactive system. Each scenario and anti-scenario is specified as a sequential thread of execution that isolates a specific aspect of the system behavior, desirably an individual requirement. Thus, it is also called a \emph{b-thread}. An application-agnostic execution mechanism repeatedly collects these scenarios, chooses actions that are consistent with all the scenarios, executes them, and continuously informs them of each selection. Previous demonstrations of BP include a show case of a fully functional nano-satellite~\cite{BarSinai2019Satellite}, a controller for a RoboSoccer player~\cite{Elyasaf2019UsingBehavioral}, an autonomous rover~\cite{Katz2019Rover}, a reactive IoT building~\cite{Elyasaf2018LSC4IoT}, and more. We elaborate on BP in~\autoref{sec:bp}.

In many cases, behaviors, or requirements, are bound to a \emph{context}. In Chess for example, some behaviors are only relevant during a check, others are only relevant for pawns, etc. Behaviors may also interact with the context by querying or changing it. In Chess for example, one of the behaviors is to trigger the `check' context whenever an opponent pieces is threatening the king. Specifying context-aware requirements (i.e., context-dependent requirements and requirements that change the system context) with b-threads --- requires explicit, \emph{first-class citizen} idioms for referencing and changing the system context --- something that is not defined for BP. In~\cite{Elyasaf2018LSC4IoT} and~\cite{Elyasaf2019UsingBehavioral}, context idioms were proposed for two implementations of BP, in live-sequence charts (LSC)~\cite{Damm2001LSCsBreathing} and in JavaScript (respectively). Using the extended languages with first-class citizen context idioms, they demonstrated how the new idioms allow for a direct specification of context-aware requirements, resulting in a better alignment between the requirements and the specification. In both languages, the context idioms were defined as syntactic sugars on top of the original language idioms, and translational semantics were proposed for the new idioms. 

In this paper, we generalize the approach of~\cite{Elyasaf2018LSC4IoT, Elyasaf2019UsingBehavioral} and present \emph{Context-Oriented Behavioral Programming} (COBP) --- a novel, language-independent paradigm for developing context-aware systems, centered on natural and incremental specification of context-dependent behaviors. Specifically, we propose to add first-class citizen context idioms to BP and define new formal semantics for specifying the relation between the system context and the system behavior. One of the advantages of these formal semantics over the translational semantics of~\cite{Elyasaf2018LSC4IoT, Elyasaf2019UsingBehavioral}, is the ability to directly implement the paradigm in different programming languages, rather than relying on existing BP implementations and translating to them. Based on this ability, we also present a JavaScript-based implementation for the paradigm and provide two case-study systems that were developed with it (Sections~\ref{sec:case-studies:motivation} to~\ref{sec:case-studies:iot}).

Another approach for integrating context with programming is the~\emph{context-oriented programming} (COP) paradigm \cite{Costanza2005LanguageConstructs}. Over the last decade, COP has evolved in a variety of languages and approaches, starting from Costanza and Hirschfeld~\cite{Costanza2005LanguageConstructs}. While there are many variations in the way they handle the contextual data and the relevant behavioral variations, the \emph{layers} is the most widespread model by far~\cite{Salvaneschi2012ContextOrientedProgramming}. Layers are a language abstraction, grouping definitions of partial methods that implement some fragment of an aspect of the system behavior~\cite{Costanza2005LanguageConstructs}. We compare some of the idioms of COP and COBP in~\autoref{sec:case-studies:ros}.

Both in COP and COBP, the system may have two contradicting behaviors, as long as they are bound to different contexts. Consider for example conflicting requirements, overridden by the context, like ``vacuum the carpet'' and ``do not vacuum while someone is asleep''. Composing the contexts and the behaviors at runtime, may lead to unpredictable behavior. Thus, both context-oriented approaches require reasoning and formal verification techniques for verifying their software, i.e., that it will function correctly in all contexts and combinations thereof. Since reasoning algorithms depend on having a formal specification of the system, an effort has been made in developing formal semantics for COP (as elaborated in~\autoref{sec:related-work}).
For this reason, one of the major purposes of this paper is to generalize the translational semantics of~\cite{Elyasaf2018LSC4IoT, Elyasaf2019UsingBehavioral}, and define formal semantics for the COBP paradigm. As we demonstrate in~\autoref{sec:case-studies:iot:discussion}, a COBP model that is based on these semantics, allows for applying reasoning techniques on the entire system (i.e., context, behavior, and execution mechanism) with no further input needed, as opposed to some COP implementations that allow for applying reasoning techniques only for part of the model, or require a manual translation of the code into a formal model. 

Furthermore, one of the key advantages of scenario-based programming (SBP)~\cite{Damm2001LSCsBreathing} in general, and of BP in particular, is the amenability of the software artifacts to formal analysis and synthesis. As we elaborate in~\autoref{sec:related-work}, most of the tools for BP and SBP rely on the mathematically rigorous nature of the semantics in providing tools for running formal analysis and synthesis algorithms. The addition of context improves the modularity of the specification, thus may contribute each of the methods, as we demonstrate in~\autoref{sec:case-studies:iot:discussion}. However, it requires to adapt these approaches, since they are all designed under the assumption that the only protocol between the b-threads is requesting, blocking, and triggering of events. Therefore, the formal semantics presented here lay the foundations towards such adaptations.

\paragraph{Outline}~\autoref{sec:bp} elaborates on the behavioral programming paradigm and describes its shortcoming when it comes to handling the system context. \autoref{sec:cobp} formally defines COBP, giving abstract semantics for the language and presents two implementations of the paradigm. Sections~\ref{sec:examples-preface} to~\ref{sec:examples-life} demonstrate how the abstract semantics can be used for formally specifying context-aware systems. In sections~\ref{sec:case-studies:motivation} to~\ref{sec:case-studies:iot} we provide two case studies of context-aware systems that were programmed using this implementation. In~\autoref{sec:related-work} we discuss reasoning approaches for COP and BP, as well as other approaches for integrating context with modeling and programming. We conclude this paper in~\autoref{sec:concluding}, with a discussion about future research directions.

\section[The Context of this Paper --- Behavioral Programming]{The Context of this Paper ---\\Behavioral Programming}
\label{sec:bp}
\begin{figure*}[th]
    \centering
    \begin{tikzpicture}
    [ my lts style, node distance=3.1cm ]
        \node[state,initial] (cold1) {$R=\{\mathit{Cold}\}$\\$B=\emptyset$};
        \node[state] (cold2) [right of=cold1] {$R=\{\mathit{Cold}\}$\\$B=\emptyset$};
        \node[state] (cold3) [right of=cold2] {$R=\{\mathit{Cold}\}$\\$B=\emptyset$};
        \node[state] (cold4) [right of=cold3] {$R=\emptyset$\\$B=\emptyset$};
        \node[state,initial] (hot1) [below=7mm of cold1] {$R=\{\mathit{Hot}\}$\\$B=\emptyset$};
        \node[state] (hot2) [right of=hot1] {$R=\{\mathit{Hot}\}$\\$B=\emptyset$};
        \node[state] (hot3) [right of=hot2] {$R=\{\mathit{Hot}\}$\\$B=\emptyset$};
        \node[state] (hot4) [right of=hot3] {$R=\emptyset$\\$B=\emptyset$};
        
        \path 
         (cold1) edge[loop above] node{otherwise} (cold1)
         (cold2) edge[loop above] node{otherwise} (cold2)
         (cold3) edge[loop above] node{otherwise} (cold3)
         (hot1) edge[loop below] node{otherwise} (hot1)
         (hot2) edge[loop below] node{otherwise} (hot2)
         (hot3) edge[loop below] node{otherwise} (hot3)
         
         (cold1) edge node {$\mathit{Cold}$} (cold2)
         (cold2) edge node {$\mathit{Cold}$} (cold3)
         (cold3) edge node {$\mathit{Cold}$} (cold4)
         (hot1) edge node {$\mathit{Hot}$} (hot2)
         (hot2) edge node {$\mathit{Hot}$} (hot3)
         (hot3) edge node {$\mathit{Hot}$} (hot4);
    \end{tikzpicture}
  \caption{A b-program that pour `Cold' and `Hot' water three times each, that consists of a single b-thread for each requirement. The b-threads are executed simultaneously, requesting, and blocking events at each of their state (the $R$ and $B$ sets).}
  \label{fig:hot-cold-base}
\end{figure*}
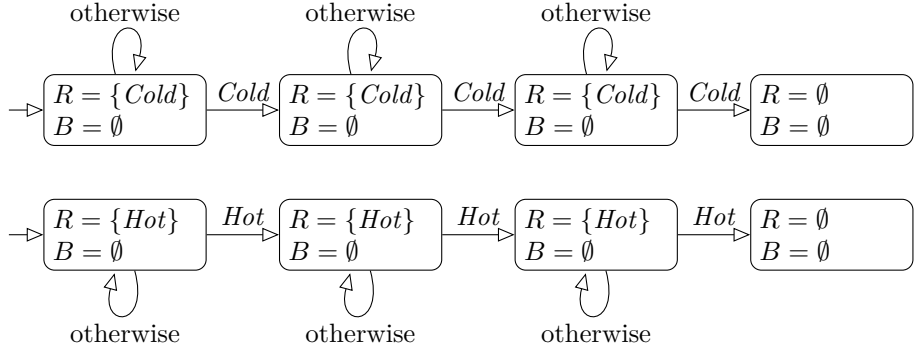

When creating a system using BP, developers specify a set of scenarios that may, must, or must not happen. Each scenario is a simple sequential thread of execution, called \emph{b-thread}, that is normally aligned with a system requirement, such as ``stop moving'' or ``turn to target''.

The set of b-threads in a model is called a behavioral program (\emph{b-program}). During run-time, all b-threads participating in a b-program are combined, yielding a complex behavior that is consistent with all said b-threads. As we elaborate below, unlike other paradigms, such as functional programming or object-oriented programming, BP does not force developers to pick a single behavior for the system to use. Rather, the system is allowed to choose any compliant behavior. This allows the run-time to optimize program execution at any given moment, e.g., based on available resources. The fact that all possible system behaviors comply with the b-threads (and thus with the system requirements), ensures that whichever behavior is chosen, the system will perform as specified. 

To make these concepts more concrete, we now turn to a tutorial example of a simple b-program, first presented in~\cite{Harel2012BehavioralProgramming}. All of the examples in this paper are specified formally, based on the transition systems that we define in~\autoref{sec:cobp}. JavaScript-based programs are presented in sections~\ref{sec:case-studies:motivation} to~\ref{sec:case-studies:iot}.

Consider a system that controls taps of hot and cold water, whose output flows are mixed, with the following requirements:
\begin{enumerate}
\item When the system loads, pour some small amount of \emph{cold} water three times.
\item When the system loads, pour some small amount of \emph{hot} water three times.
\end{enumerate}
\autoref{fig:hot-cold-base} shows a b-program that fulfills these requirements. It consists of two b-threads, added at the program start-up. The first is responsible for fulfilling requirement \#1, and the second fulfills requirement \#2.

\begin{figure}[t]
  \centering
  \begin{tikzpicture}
    \pic {bpcycle};
  \end{tikzpicture}
  \caption{The life cycle of a b-program (adapted from~\cite{Harel2012BehavioralProgramming}). B-threads repeatedly execute an internal logic and then synchronize with each other, by submitting a synchronization statement to a central event arbiter. Once all b-threads have submitted their statements, the central event arbiter selects an event that was requested and was not blocked. B-threads that either requested or waited for this event are resumed, while the rest remain paused for the next cycle.}
  \label{fig:bp-lifecycle}
\end{figure}
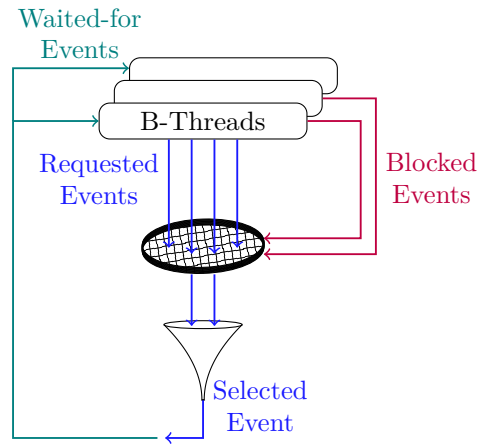

The program's structure is aligned with the system requirements, with a single b-thread for each requirement, requesting, and blocking events at each of their state (the $R$ and $B$ sets). Harel, Marron and Weiss~\cite{Harel2010ProgrammingCoordinated} proposed a simple protocol for interweaving the b-threads and executing the model, depicted in \autoref{fig:bp-lifecycle}. B-threads repeatedly execute an internal logic and then synchronize with each other, by submitting a synchronization statement to a central event arbiter. Once all b-threads have submitted their statements, the central event arbiter selects an event that was requested and was not blocked. B-threads that either requested or waited for this event (specified on the edges in our example) are resumed, while the rest of the b-threads remain paused for the next cycle. Back to our example, the specification in~\autoref{fig:hot-cold-base} does not dictate an order in which actions are performed since the b-threads do not block events of each other. Thus, any of the following runs are possible: Cold-Cold-Hot-Hot-Cold-Hot, or Cold-Hot-Cold-Hot-Cold-Hot, etc. This contrast with, say, imperative programming languages (e.g., Java, C) that would have to dictate exactly when each action should be performed. Thus, traditional programming paradigms are prone to over specification, while behavioral programming avoids it.

While a specific order of actions was not required originally, this  behavior may represent a problem. Consider for example an additional requirement that the client requested after running the initial version of the system:
\begin{enumerate}
\setcounter{enumi}{2}
\item Two actions of the same type cannot be executed consecutively.
\end{enumerate}

\begin{figure}[t]
    \centering
  \begin{tikzpicture}[my lts style]
        \node[state,initial] (n1) {$R=\emptyset$\\$B=\{\mathit{Hot}\}$};
        \node[state,initial] (n2) [right of=n1] {$R=\emptyset$\\$B=\{\mathit{Cold}\}$};

        \path 
         (n1) edge [bend left, above] node {$\mathit{Cold}$} (n2)
         (n2) edge [bend left, below] node {$\mathit{Hot}$} (n1);
    \end{tikzpicture}
  \caption{The \texttt{Interleave} b-thread --- ensures that two actions of the same type cannot be executed consecutively, by blocking an additional request of `Cold' until the `Hot' is performed, and vice-versa.}
  \label{fig:hot-cold-interleave}
\end{figure}
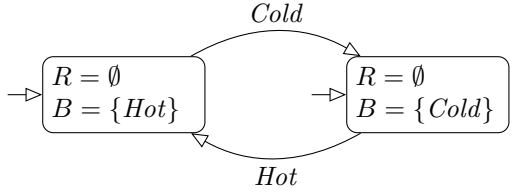

While we may add a condition before requesting `Cold' and `Hot', the BP paradigm encourages us to add a new b-thread for each new requirement. Thus, we add a third b-thread, called \texttt{Interleave}, presented in \autoref{fig:hot-cold-interleave}.

The \texttt{Interleave} b-thread ensures that there are no repetitions, by forcing an interleaved execution of the performed actions: `Hot' is blocked until `Cold' is executed, and then `Cold' is blocked until `Hot' is executed. Note that this b-thread can be added and removed without affecting other b-threads. This is an example of a \emph{purely additive} change, where the system behavior is altered to match a new requirement without affecting the existing behaviors. While not all changes to a b-program are purely additive, many useful changes are, as demonstrated in~\cite{Harel2010ProgrammingCoordinated}. 

\subsection{The Absence of Context in BP}
\label{sec:bp:drawback}
In this paper, the term context refers to information that can be used to characterize the situation of system's entities or processes. The term \emph{`context awareness'} refers to the system's ability to use contextual information~\cite{Dey1999BetterUnderstanding}. Examples for context-aware systems are, adaptive system, ubiquitous systems, and mobile systems, as they all require the ability to change behavior with respect to context. 

Consider for instance, a context-aware version of the above hot-cold example, with many rooms and taps. The system may have context-dependent requirements that define different behaviors, depending on the context, e.g., the type of the room, the person who uses the taps, etc. While it is possible to express these aspects without explicit reference to context, we will show that direct idioms for working with context allow to reuse the scenarios and to make them more aligned with how requirements are defined (which was the main design goal for BP). 
 

All the implementations of BP that we are aware of, do not support direct communication between b-threads (with an exception of LSC that we discuss in~\autoref{sec:related-work}). On the contrary, as elaborated in~\autoref{sec:cobp:semantics:bp}, the formal semantics of BP define that the only data shared among the b-threads --- are the events. Yet, sharing the system context among the b-threads, overrides this definition. Thus, in practice, describing context-aware systems using BP forces the programmer to use some ``hacks'', such as sharing data using global variables, which violates the formal definition and breaks the ability to formally verify the programs. Another ``hack'' is to send the entire system context embedded in each event, making a lot of redundant code and breaking the alignment between the b-threads and the requirements.

\section{Context-Oriented Behavioral Programming}
\label{sec:cobp}
To address this problem, we present \emph{Context-Oriented Behavioral Programming (COBP)} --- an extension to the behavioral-programming paradigm with explicit idioms for defining and referencing the context. As previously noted, this approach was first presented in~\cite{Elyasaf2018LSC4IoT}, where the Live Sequence Charts (LSC) language was extended with context idioms, along with a methodology for developing context-aware systems using the extended language. In~\cite{Elyasaf2019UsingBehavioral}, a similar extension has been made to BPjs --- an environment for running behavioral programs written in JavaScript~\cite{BarSinai2018BPjs}. In both cases, the semantics were defined by translating the context idioms to the existing idioms of LSC and BPjs. 
In this paper we generalize these extended languages by defining abstract semantics (rather than translational semantics) and lay the foundations for further research on formal analysis and synthesis. 

In COBP, the system state incorporates the context of the system that can be queried and updated. For example, in the context-aware version of the hot-cold system, we may define the context to be the entire building, including the rooms, the taps, the amount of hot/cold units that need to be poured, and any additional required data. We will also define queries on the context such as $room\,with\,taps$ and $\mathit{kitchen}$, that return a list of all rooms with taps and of all kitchens, respectively. A b-thread in COBP is a context-aware b-thread, denoted as \emph{CBT}, that is bound to a certain query on the context. Whenever there is a new result to the query, a new \emph{live copy} of the CBT is spawned, with the query result given as a local variable to the live copy. Thus, we refer to this result as the \emph{seed} of the live copy. Similar to a b-thread, a CBT also specifies an aspect of the system behavior that is relevant to each new seed that is generated, allowing to use its data, and to query and update the context. Back to our extended hot-cold system, we can convert the original three b-threads into CBTs and bind them to the appropriate query (i.e., $room\,with\,taps$). Thus, the queries represent both the context of the requirements and the context of the CBTs. The complete example of the extended hot-cold system is given in~\autoref{sec:examples-ext-hot-cold}.

\begin{figure}[t]
  \centering
  \begin{tikzpicture}
    \pic {cobpcycle};
  \end{tikzpicture}
  \caption{The life cycle of a context-aware b-program. Each context-aware b-thread (CBT) is bound to a query on the contextual data. Whenever a new answer exists for a query, new live copies are spawned for the relevant CBTs. The live copies repeatedly execute an internal logic that may depend on the contextual data and then synchronize with each other, by submitting a synchronization statement to a central event arbiter. Once all live copies have submitted their statements, the arbiter selects an event that was requested and was not blocked. The event is passed to the \emph{Effect Function} that may update the contextual data, depending on its specification. The (updated) contextual dataset is passed back to the CBTs, along with the selected event. All CBTs that either requested or waited for this event are resumed, while the rest remain paused for the next cycle.}
  \label{fig:cobp-lifecycle}
\end{figure}
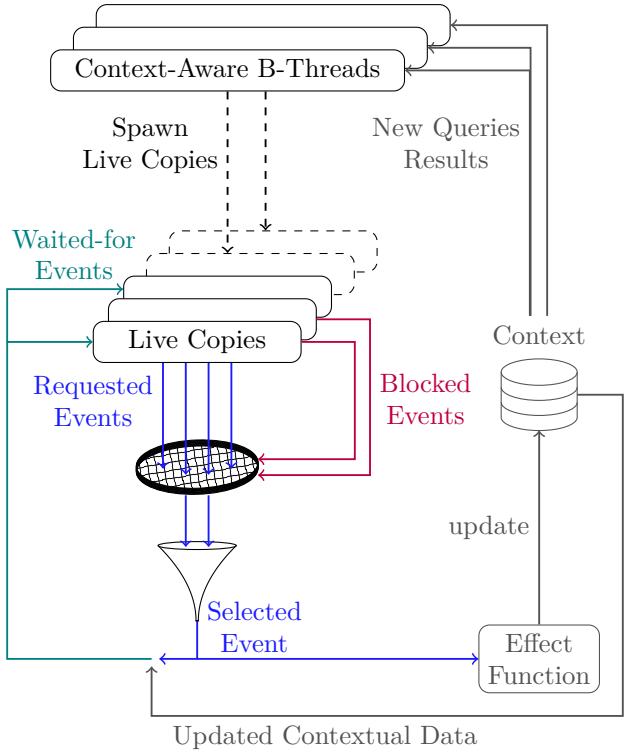

The life cycle of COBP contains the life cycle of BP with several modifications, as depicted in~\autoref{fig:cobp-lifecycle}:
\begin{itemize}
    \item The live copies are the ones that are executed, and the CBTs spawn new live copies upon each new answer to their query. 
    \item The internal logic of the live copies does not only depend on the last event, as in BP, but also on the context.
    \item An effect function receives the selected event and updates the contextual data according to its specification. Then, the selected event and the updated context are transferred back to the live copies.
\end{itemize}

We now turn to define the abstract semantics of COBP. We begin with the BP semantics, adopted from the definitions of~\cite{Harel2010ProgrammingCoordinated}, since the COBP semantics extend them. 

\subsection{BP Semantics}
\label{sec:cobp:semantics:bp}
The definition outline of the BP semantics is as follows: The definition of a labeled transition system (LTS) is given in~\autoref{def:transition-system}. In~\autoref{def:bt}, we define a b-thread as a labeled transition system in which events at each state can be marked as requested or as blocked.  Finally, \autoref{def:bt-run}, defines the two basic rules for executing a set of b-threads: (1) An event occurs if and only if it is requested by some b-thread and is not blocked by any b-thread; (2) All b-threads affected by a given event undergo a state transition when the event occurs.

\begin{definition}[labeled transition system~\cite{Keller1976FormalVerification}]
\label{def:transition-system}
A labeled transition system is a quadruple $\langle S, E, \rightarrow, init \rangle$, where $S$ is a set of states, $E$ is a set of events, $\rightarrow$ is a subset of $S \times E \times S$ called a transition relation, and $init \in S$ is an initial state. The runs of such a transition system are sequences of the form $s_0 \xrightarrow{e_1} s_1 \xrightarrow{e_2} \cdots \xrightarrow{e_i} s_i \cdots$, where $s_0 = init$, and for all $i = 1, 2, \cdots , s_i \in S, e_i \in E$, and $ \langle  s_{i-1},e_i,s_i \rangle \in \to$ (this membership will, from now, be written as $s_{i-1} \xrightarrow{e_i} s_i$ for shortening and for simplifying the notions).
\end{definition}

\begin{definition}[behavior thread]
\label{def:bt}
A behavior thread over an event-set $E$ is a tuple $\langle S, \rightarrow, init, R, B \rangle$, where $\langle S, E,\allowbreak \rightarrow, init \rangle$ forms a labeled transition system, $R: S \rightarrow 2^E$ is a function that associates each state with the set of events requested by the b-thread when in that state, and $B: S \rightarrow 2^E$ is a function that associates each state with the set of events blocked by the b-thread when in that state.
\end{definition}

\begin{definition} [runs of a set of b-threads]
\label{def:bt-run}
The runs of a set of b-threads $\{\langle S_i, \rightarrow _i, init_i, R_i, B_i\rangle\}_{i=1}^n$ , all over the event-set $E$, is defined as the runs of the labeled transition system $\langle S, E, \rightarrow, init \rangle$, where $S = S_1 \times \cdots \times S_n$, $init = \langle init_1, \cdots , init_n \rangle$, and $\rightarrow$ includes a transition $\langle s_1,\cdots,s_n \rangle \xrightarrow{e} \langle s'_ 1,\cdots,s'_n \rangle$ if and only if\vspace{1ex}

\begin{tabular}{lc@{\hspace{10mm}}l}
     \phantom{and }& $ \underbrace{e \in \bigcup \limits_{i=1}^n R_i(s_i)}_{\text{e is requested}} \;\;\; \bigwedge \;\;\; \underbrace{e \notin \bigcup \limits_{i=1}^n B_i(s_i)}_{\text{e is not blocked}} $ & $(1)$ \\
\multicolumn{3}{l}{and} \\ 
     & $\underbrace{\bigwedge\limits_{i=1}^n (s_i \xrightarrow{e} s'_i)}_{\text{b-threads react to events}}$ & $(2)$
\end{tabular}\vspace{1ex}
\end{definition}

Note that for the sake of simplifying the presentation, our definitions are very similar to those presented in~\cite{Harel2010ProgrammingCoordinated}, but not identical. The difference is that~\cite{Harel2010ProgrammingCoordinated} assigned a separate set $E_i$ to each b-thread and defined the joint event-set as $E = \bigcup _{i=1}^n E_i$.  With this, they could automatically add an implicit self-loop at each b-thread's state for any $e \notin E_i$ (i.e., event that the b-thread is not aware of). Here, to simplify the definitions, we explicitly add to all of our examples an ``otherwise" self-loop to all b-thread states. 


\subsection{COBP Semantics}

We begin with the definition of a context-aware behavioral program and continue with its execution semantics. 

\subsubsection{Context-Aware Behavioral Program} 
A context-aware behavioral program is composed of the context data structure with its management functions (definitions~\ref{def:context} and~\ref{def:context:effect}), together with a set of context-aware b-threads (\autoref{def:context:cbt}). A context-aware b-thread (CBT) is a behavioral thread that is bound to a certain query on the context. Our definition for context is inspired by the definition of~\cite{ Dey1999BetterUnderstanding} --- information that can be used to characterize the situation of entities or processes in a system. 

\begin{definition}[context, queries, and updates]
\label{def:context}
$\mathit{CTX}$\\is a (possibly infinite) set of all possible contexts that the system is designed to be aware of. At any given moment the system is in a context $\mathit{ctx} \in \mathit{CTX}$. The management of the context is done by the sets $\mathit{QUERY}$ and $\mathit{UPDATE}$, where each member $\mathit{query} \in \mathit{QUERY}$ is a function with $\mathit{dom}(\mathit{query})=\mathit{CTX}$ (and arbitrary range) and each member $\mathit{update} \in \mathit{UPDATE}$ is a function from $\mathit{CTX}$ to $\mathit{CTX}$. Finally, $\mathit{ctx}_\mathit{init}$ denotes the initial context of the system. 
\end{definition}

As said before, we envision a complex data structure representing the different aspects of the contextual data. The set $CTX$ is a mathematical representation of all possible states of this data structure. While it may read as if we only allow one context at a time for the whole system, this context is actually a combination of many contextual aspects, as we demonstrate in the examples below.

In applications, we envision a language for querying and updating the context state such as SQL or procedural code, as we will show. In this case, the result of a query can be, for example, a tuple in a table or the number of people in a room. We therefore do not to restrict the range of the $query$ function in our mathematical abstraction. 

\begin{definition}[the effect function]
\label{def:context:effect}
The connection between the events that the system generates and the updates of the context are defined by the function $\mathit{effect}\colon E \to \mathit{UPDATE}$ that maps events to context update functions.
\end{definition}

Note that we separate the definition of the event from the definition of its effect to enforce a separation of concerns between behavioral aspects and data aspects.

\begin{definition}[a context-aware behavior thread]
\label{def:context:cbt}
A context-aware behavior thread (cbt) over an event-set $E$ is a tuple $\langle S, \rightarrow, \mathit{init}, R, B, \mathit{query}\rangle$, where $S, E,$ and $\mathit{init}$ are as defined in \autoref{def:bt}, $\mathit{query} \in \mathit{QUERY}$ is a query that the cbt is bound to. Here, the transition relation $\rightarrow$ is a subset of $S \times \mathit{CTX} \times \mathit{range}(\mathit{query}) \times E \times S$, meaning that the transition of a contextual b-thread may be conditioned not only on the last event, but also on the context at the time of the transition and on a query result. Similarly, $R$ and $B$ can be conditioned, i.e., they are members of the set $S \times \mathit{CTX} \times \mathit{range}(\mathit{query}) \rightarrow 2^E$. To simplify future notations, we define the functions $\mathit{init}(\mathit{cbt}) = \mathit{init}$, and $\mathit{query}(\mathit{cbt}) = \mathit{query}$.
\end{definition}

Conceptually, CBTs play the same role as regular b-threads. The key addition is the ability to explicitly connect the behavior of the b-thread with the context and with the answer of the query that seeded the b-thread. For example, in Chess, one can directly specify the castling requirement: ``(1) Neither king nor rook involved in castling may have moved from the original position; (2) There must be no pieces between the king and the rook.''\footnote{\url{www.chesscoachonline.com/chess-articles/chess-rules}.}

\subsubsection{Execution Semantics for a Context-Aware Behavioral Program}
In BP, the b-threads are executed, however in COBP, CBTs are not directly executed. Whenever an aspect of the context becomes active (i.e., there is a new result to the query) --- a new live copy (\autoref{def:context:lc}) of each of the CBTs that are bound to the context --- is spawned, with the result of the query used as the seed of the live copy, passed to it as a local variable. The execution semantics of a context-aware behavioral program (\autoref{def:context:run}) define a new LTS, where its states are composed of the active live copies and the state of the system context. The transition function defines how events are selected, how the context is affected by the selected event, and how live copies are spawned based on changes of the system context.

\begin{definition}[a live copy of a context-aware behavior thread]
\label{def:context:lc}
A live copy (instance) of a context-aware behavior thread is a tuple $\langle \mathit{cbt}, s, c \rangle$ where $\mathit{cbt}=\langle S, \rightarrow,\allowbreak \mathit{init}, R, B, \mathit{query}\rangle$ is a context-aware behavior thread, $s\in S$, and $c\in \mathit{range(query)}$. We refer to $c$ as the seed of the live copy. To simplify the notations, we extend the definition of $R$ and $B$ (see~\autoref{def:bt}) to accept a live copy: $R(\langle \mathit{cbt}, s, c \rangle) = R(s)$ and  $B(\langle \mathit{cbt}, s, c \rangle) = B(s)$. 
\end{definition}

\begin{definition}[runs of context-aware b-threads]
\label{def:context:run}
The runs of a set of context-aware b-threads over an event-set $E$, and a cbt-set $\mathit{CBT}$, is defined as the runs of the LTS $\langle S, E, \rightarrow, init \rangle$, where:

\begin{enumerate}[leftmargin=*]
\item \scaletoline{$S = \{ \langle \mathit{ctx}, \mathit{LC} \rangle \colon \mathit{ctx} \in \mathit{CTX}, \mathit{LC} \text{ is a set of live copies}\}$}

\item $\mathit{init} = \langle \mathit{ctx}_{init}, \{ \langle \mathit{cbt}, s, c \rangle \colon \mathit{cbt} \in \mathit{CBT}, s = \mathit{init}(\mathit{cbt}),\allowbreak c \in \mathit{query}(\mathit{cbt})(\mathit{ctx}_{\mathit{init}}) \}\rangle$

\item $\rightarrow$ includes a transition $\langle \mathit{ctx}, \mathit{LC} \rangle \xrightarrow{e} \langle \mathit{ctx}', \mathit{LC}' \rangle$ if and only if
\begin{enumerate}[leftmargin=1pt] \setlength{\itemsep}{10pt}%
        \item $ \underbrace{e \in \bigcup \limits_{\mathit{lc} \in \mathit{LC}} R(lc)}_{e \text{ is requested}} \;\;\; \bigwedge \;\;\; \underbrace{e \notin \bigcup \limits_{\mathit{lc} \in \mathit{LC}} B(\mathit{lc})}_{e \text{ is not blocked}} $ 
        
        \item $\mathit{ctx}' = \mathit{effect}(e)(\mathit{ctx})$
        
        \item $\mathit{LC}'=  \mathit{LC}'_\mathit{run} \cup \mathit{LC}'_\mathit{new}$ where 
        
        \vspace{1em}\hspace{-20pt}
        $\underbrace{LC'_{run}=\{  \langle cbt,s',c\rangle  \colon  \langle cbt, s, c\rangle \in LC \text{ and } s \xrightarrow{e,\, ctx',\, c}_{cbt} s' \}}_{\text{running live copies (based on transition systems)}}$
        
        and 
        
        \hspace{-20pt}$\underbrace{LC'_{new}= \left\{ 
        \begin{array}{l@{\hspace{1em}}l}
             \multicolumn{2}{l}{\langle cbt, init(cbt), c \rangle \colon cbt \in CBT,} \\
             & c \in query(cbt)(ctx')\setminus query(cbt)(ctx)
        \end{array}
        \right\}}_{\text{spawned live copies (based on new query results)}}$.
    \end{enumerate}

\end{enumerate}
\end{definition}

\subsection{A Short Discussion}
Compared to the BP semantics, these semantics embed the state of the context in the system state and extend the transition function to support the new state. Since the semantics formally specify the entire system (i.e., context, behavior, and execution mechanism) --- reasoning techniques can be applied on the complete system/model with no further input (as elaborated and demonstrated in~\autoref{sec:case-studies:iot:discussion}). Moreover, the formal semantics, allow for generating different perspectives of the model, such as:
\begin{itemize}
    \item \textbf{Context-behavior perspective:} Represents relations between contexts and behaviors, such as: ``which behaviors are bound to a certain context'', ``affected by the termination of a certain context'', etc.
    \item \textbf{Behavior-context perspective:} Represents how behaviors affect the context, e.g., which contexts queries can be affected by a certain CBT.
    \item \textbf{Context-context perspective:} Represents the relations between different aspects of the context, such as: temporal (e.g., contexts $\mathit{day}$ and $\mathit{night}$ cannot be active simultaneously) and structural (e.g., the $\mathit{building}$ context is composed of $\mathit{room}$s, a $\mathit{kitchen}$ instance is also a $\mathit{room}$ instance, etc.).
\end{itemize}

These perspectives (and others), may ease the development process by improving the readability, explainability, and correctness of the model. 

Another interesting feature of the semantics is the process of \emph{context activation}, i.e., adding a result to a query and spawning live copies. We note that only $\mathit{UPDATE}$ functions may change the context, and they are triggered only by the $\mathit{EFFECT}$ function, that is triggered only upon selecting an event. Thus, while live copies are advanced simultaneously, only one $\mathit{UPDATE}$ function can be triggered during a system transition. Changes to the context may yield new results for one or more $\mathit{QUERY}$ functions and upon such a change, all the CBTs that are bound to these queries will spawn live copies simultaneously. For example, once a new room of type kitchen is added to the context (either dynamically by using the effect function mechanism or statically by adding it to the initial context), all CBTs that are bound to the $room\,with\,taps$ query and $\mathit{kitchen}$ query spawn live copies simultaneously. This behavior is desired if there is no special requirement for an order. If there is such a requirement, then the BP and COBP paradigms encourage us to specify this additional requirement with an additional CBT (like the third, interleave b-thread of the hot-cold example). Adhering to the formal semantics, our JavaScript implementation delicately deals these nuances and others (elaborated below).

The implementation of the event arbitrer, that continuously selects a requested and not blocked event, can be either na\"ive, i.e., select one randomly, or advanced, e.g., by using priorities, heuristics, learning, etc. Again, this behavior is desired assuming that there is no special requirement for ordering the events. Of course, such a degree of parallelism may cause unpredictable behavior, and while the specification may be aligned to the requirements, there may be problems with the requirements themselves. In order to deal with this problem, different reasoning approaches have been proposed for BP, as we elaborate in~\autoref{sec:related-work}. We further demonstrate the process of context activation in the examples below.

\subsection{Implementations}
\label{sec:cobp:implementation}
In this section, we describe two COBP implementations and discuss their relation to the semantics. The first one is based on the PlayGo implementation of LSC~\cite{Harel2010PlayGoComprehensive} and is fully described in~\cite{Elyasaf2018LSC4IoT}. The second implementation~\cite{Elyasaf2019UsingBehavioral} is based on a JavaScript implementation of BP, called BPjs~\cite{BarSinai2018BPjs}. In both implementations, context idioms (i.e., CBT, effect function, etc.) are translated into idioms of the underline BP implementation.

Both the PlayGo and the BPjs tools allow for scenarios (or b-threads) to share data objects between them. Yet, to avoid a violation of the BP semantics, it is ``well known'' among BP developers that scenarios are prohibited to indirectly alter the behavior of each other by modifying shared objects. The COBP semantics address this issue by formalizing the connection between the data objects and the behavioral specification. According to~\autoref{def:context:run}, the change to $ctx$ during a transition (i.e.,~\autoref{def:context:run} [3.b]) must occur before advancing the set of live copies (i.e.,~\autoref{def:context:run} [3.c]). Our implementations use shared objects for maintaining the system context (i.e., $ctx$), and synchronize the access using the BP execution mechanism that chooses an event and advances the b-threads. 

\newsavebox{\tempbox}
\sbox{\tempbox}{
    \begin{LSCcontextualdiagram}[title width=p{120px}]{Turn the light on/off based on the room occupancy}{r $\in$ NonemptyRoom}{}
      \newlifeline[edge distance=1,type=dynamic]{light}{r.light}
      \newlifeline[edge distance=1.4,type=dynamic]{room}{r}
      \coldexecute{room}{on()}{light}
      \coldmonitor{room}{ended(``Nonempty room'')}{room}
      \coldexecute{room}{off()}{light}
    \end{LSCcontextualdiagram}
}

\newbox{\mybox}
\begin{lrbox}{\mybox}
\begin{lstlisting}[linewidth=8.6cm]
bp.registerCBT("Turn the light on/off based on " +
  "the room occupancy", 
  "NonemptyRoom", // the query
  function(r) {
    bp.sync({ request: on(r.light) });
    bp.sync({ waitFor: CTX.Ended("NonemptyRoom", r) });
    bp.sync({ request: off(r.light) });
  });
\end{lstlisting}
\end{lrbox}

\begin{figure*}[t]
    \centering
    \subfloat[The CBT in LSC. Each vertical line, called a lifeline, represents a data object and the arrows define the events. The solid arrows represent a request, and waiting for an event is represented by the dashed arrow. 
    \label{fig:implementations:lsc}]
    {\adjustbox{valign=b,margin=2ex 0ex}{\usebox\tempbox}}
    \qquad\quad
	\subfloat[The CBT in JavaScript. The query name is defined in the second parameter of the \lstinline|registerCBT| function. 
	\label{fig:implementations:bp}]
	{\adjustbox{valign=b}{\usebox\mybox}}
    \caption{A CBT (in LSC and JavaScript) that specifies that a room's light should be turned on/off based on its occupancy. Since the scenario is bound to the \lstinline|NonemptyRoom| query (defined in the DAL), once a room (any room) becomes occupied, a new live copy of the scenario is spawned with the room given as the local variable $r$.}
    \label{fig:implementations}
\end{figure*}

\subsubsection{System Design}
To increase the modularity and the separation of concerns of specifications, we adapt the \emph{multilayered architectural pattern}. We note that according to the BOCP semantics, the CBTs are unaware of the context schema, the $\mathit{UPDATE}$ set, how each $q \in \mathit{QUERY}$ is defined, and how the effect function is defined. This information represents data-related aspects of the system, rather than behavioral, and the interface between the CBTs and the context is defined by the queries' names and the contract of the effect function (i.e., the effect of each event). With respect to these insights, we define the layers as follows:
\begin{enumerate}
\item The \emph{business-logic layer} (BLL) consists of the behavioral specification --- the set of CBTs that specify the behavior of the system. When developing this layer, the programmer must be aware of which information is accessible from the context (i.e., the queries' names and the structure of their results) and how events affect the context.

\item The \emph{data-access layer} (DAL) abstracts the context-related decisions for the BLL. The contextual data is stored in a structured dataset (e.g., a relational database, classes, etc.) and managed using a data-access layer (DAL). The DAL is defined by a \emph{``query and command repository''} and an effect function. The ``query and command repository'' manages the contextual data using a high-level query language (e.g., SQL, methods, etc.). The context's `select' queries allow for triggering a stored procedure whenever a record is added or removed from its result (in analogy to ``database views''), and the context's `update' commands update the data as required (by adding, deleting, and changing objects and object relations). Based on the selected event, the effect function uses the `update' commands for changing the context. To simplify the notations of the examples below, we define the effect of all events as the identity function (i.e., they do not change the context), unless explicitly specified otherwise. 

\item The \emph{application (or service) layer} defines the interaction between the BLL and the environment (e.g., a GUI, network messages, etc.). This interaction is already defined in BP by using a publish-subscribe mechanism, where external, environmental events are selected only at a \emph{super step}, i.e., when there are no more internal events to select. Similarly, the environment may listen to internal events and act upon them (e.g., trigger actuators). The complete mechanism is elaborated in~\cite{Harel2010ProgrammingCoordinated}.
\end{enumerate}

The multilayered architecture facilitates the separation of concerns between the layers, and by binding b-threads to queries, the modularity of the behavioral specification is increased. \autoref{fig:implementations} depicts how CBTs are specified using this design. In the examples and case studies below, we follow this design and separate each implementation into three sections --- \emph{Context Specification} (for the DB schema), \emph{Data-Access Layer}, and \emph{Behavioral Specification}.

\subsubsection{Spawning an Terminating Live Copies}
PlayGo includes a mechanism for spawning live copies upon changes to data objects, allowing us to use it for dynamically activating and terminating CBTs' live copies. BPjs does not have such mechanism, though it can be created by adding a b-thread that waits for contextual-data updates and then requests an event that contains a list of new/ended results to the queries. This event is used for spawning and terminating live copies, as elaborate next. 

We implement CBT as a function that registers a regular b-thread (see \autoref{lst:implementations:cobpjs-cbt}). The b-thread waits for new-results announcement and registers another b-thread for each result (for executing $func$ in parallel), with the new answer given as a parameter.
\begin{lstlisting}[
    % float=t,
    label={lst:implementations:cobpjs-cbt},
    caption={The implementation of a CBT in BPjs.},
]
bp.registerCBT = function(name, query, func) {
  bp.registerBThread("cbt: " + name, function() {
    while(true) {
      bp.sync({ waitFor: CTX.NewResults(query)})
        .data.forEach(c => 
          bp.registerBThread("Live copy: "+name+" "+c.id,
            function() { func(c); }); );
    }}); }
\end{lstlisting}

In some cases, a live-copy should terminate whenever its seed no longer answers its query. In Chess, for example, live copies that handle a piece's behavior are no longer relevant once the piece is captured. The implementations and semantics of BP and COBP allow for b-threads/live copies to terminate themselves. Thus, live copies may listen to the aforementioned announcement event and terminate themselves. This behavior can be either explicitly specified for certain states, or it can be implicitly applied to all. While the latter may be preferred when the ``certain states'' are all of them, caution is required since live copy may need to release resources or do some actions before terminating. 

\section{Examples --- Motivations and Overview}
\label{sec:examples-preface}
To demonstrate how context-aware systems can be specified, executed, and verified, using the abstract semantics, we now present two toy examples that are given here merely for understanding these concepts. 
For each example, we demonstrate its ``execution'' using the formal semantics, showing how the context and the live copies change over time. 

While here we implement the examples abstractly, they have a JavaScript implementation as well, that can be downloaded and executed from~\url{https://github.com/bThink-BGU/Papers-2020-COBP}. Followed by these examples, we present two real-life systems developed using our JavaScript implementation (sections~\ref{sec:case-studies:motivation} to~\ref{sec:case-studies:iot}). 

\section[Example --- The Context-Aware Hot-Cold System]{Example --- The Context-Aware Hot-Cold\\System}
\label{sec:examples-ext-hot-cold}
We now demonstrate how the extended hot-cold system can be implemented using the abstract semantics. We begin with describing the system requirements:
\begin{enumerate}
\item A building has different room types, such as: kitchen, bathroom, bedroom, living room, hall, etc.
\item Kitchens and bathrooms have a button and a hot and a cold taps.
\item For each room with taps: \begin{enumerate}
\item When the button is pressed, pour some small amount of \emph{cold} water three times.
\item When the button is pressed, pour some small amount of \emph{hot} water three times.
\end{enumerate}
\item For each kitchen --- two pouring actions of the same type cannot be executed consecutively.
\end{enumerate}

We note that requirements 3 and 4 define the desired system behavior for each room-with-taps /kitchen (respectively). Requirements 1 and 2 define data constraints on the system or data aspects of the system, thus define the schema of the system context rather than its behavior. We also note that the system context is static and does not change throughout the lifetime of this system. Of course, this is not always true as requirements may depend on dynamic changes to the context, as we demonstrate in~\autoref{sec:examples-life}. For example, a requirement may depend on personal preferences (e.g., whenever Joe uses the kitchen's taps --- pour only two units), room temperature (e.g., do not pour hot water when the room temperature is high), etc.

\subsection{Implementation}
We define the possible events to be $E=\allowbreak\{\mathit{Push_i},\mathit{Hot}_i,\allowbreak \mathit{Cold}_i \colon i \in \mathbb{N}\}$, where $i$ represents the i-th room, $\mathit{Push_i}$ represents pushing the room's button, and $\mathit{Hot_i}/\allowbreak\mathit{Cold_i}$ represents pouring hot/cold (respectively) water.

\subsubsection*{Context Specification}
\label{sec:examples-ext-hot-cold:context-spec}

\begin{itemize}[leftmargin=1em]
\item $\mathit{RoomType}=\{\mathit{kitchen},\mathit{bathroom},\mathit{living\,room}, \dots \}$ is a set of all room types.
\item The i-th room in the building is represented by $\mathit{room}_i=\langle i,\mathit{type}_i \rangle$, where $i \in \mathbb{N}$ is the room number and $\mathit{type}_i \in \mathit{RoomType}$ represents the type of the room.
\item The set of all possible system contexts, $\mathit{CTX}$, is defined as the set of all possible rooms --- $\{ \mathit{room}_i \colon i \in \mathbb{N}\}$.
\end{itemize}

\subsubsection*{Data-Access Layer}
\label{sec:examples-ext-hot-cold:dal}
As noted before, the behavioral requirements depend on two aspects of the context --- `room with taps' and `kitchen'. Therefore, the DAL has the following two queries:
\begin{itemize}
\item $\mathit{RoomWithTaps}=\{ \langle i,\mathit{type}_i \rangle \colon i \in \mathbb{N} \wedge \mathit{type}_i \in$\\$\{kitchen,\allowbreak bathroom\}\}$. 

\item $\mathit{Kitchen}=\{ \langle i,\mathit{type}_i \rangle \colon i \in \mathbb{N} \wedge \mathit{type}_i = \mathit{kitchen}\}$. 
\end{itemize}

The queries' names are used by the CBTs as an abstraction for the queries. Thus, future changes to the context specification (e.g., due to additional requirements) will not affect the definitions of the CBTs.

Since the system context is static and does not change, the effect function for all events is the identity function.

\subsubsection*{Behavioral Specification (CBTs)}
\label{sec:examples-ext-hot-cold:cbts}
We translate the b-threads of \autoref{sec:bp} to CBTs, binding the hot/cold b-threads to the $\mathit{RoomWithTaps}$ query, and the interleave b-thread to the $\mathit{kitchen}$ query:

\begin{itemize}[label={},leftmargin=0pt]
\item $\mathit{CBT}_{Cold} \colon$ 
\cbt[my cbt style,
every state/.style={
    align=left,
    text width=20mm,
    rectangle,
    rounded corners,
},
node distance=3.2cm,
]
{Cold}
{$\mathit{Query=RoomWithTaps}$}
{
    \node[state,initial] (push)[text width=10mm,right=0.5cm of desc] {$R=\emptyset$\\$B=\emptyset$};
    \node[state] (cold1)[right=1.1cm of push] {$R=\{\mathit{Cold_i}\}$\\$B=\emptyset$};
    \node[state] (cold2) [right of=cold1] {$R=\{\mathit{Cold_i}\}$\\$B=\emptyset$};
    \node[state] (cold3) [above=0.6cm of cold1] {$R=\{\mathit{Cold_i}\}$\\$B=\emptyset$};
    
    \path 
     (push) edge[loop below] node{otherwise} (push)
     (cold1) edge[loop below] node{otherwise} (cold1)
     (cold2) edge[loop below] node{otherwise} (cold2)
     (cold3) edge[loop above] node{otherwise} (cold3)
     
     (push) edge node {$\mathit{Push_i}$} (cold1)
     (cold1) edge node {$\mathit{Cold_i}$} (cold2)
     (cold2) edge[bend right,midway,right] node {$\mathit{Cold_i}$} (cold3)
     (cold3) edge[bend right,midway,left] node {$\mathit{Cold_i}$} (push);
}

\item $\mathit{CBT}_\mathit{Hot} \colon$ 
\cbt[my cbt style,
every state/.style={
    align=left,
    text width=20mm,
    rectangle,
    rounded corners,
},
node distance=3.2cm,
]
{Hot}
{$\mathit{Query=RoomWithTaps}$}
{
    \node[state,initial] (push)[text width=10mm,right=0.5cm of desc] {$R=\emptyset$\\$B=\emptyset$};
    \node[state] (hot1) [right=1.1cm of push] {$R=\{\mathit{Hot_i}\}$\\$B=\emptyset$};
    \node[state] (hot2) [right of=hot1] {$R=\{\mathit{Hot_i}\}$\\$B=\emptyset$};
    \node[state] (hot3) [above=0.6cm of hot1] {$R=\{\mathit{Hot_i}\}$\\$B=\emptyset$};
    
    \path 
     (push) edge[loop below] node{otherwise} (push)
     (hot1) edge[loop below] node{otherwise} (hot1)
     (hot2) edge[loop below] node{otherwise} (hot2)
     (hot3) edge[loop above] node{otherwise} (hot3)
     
     (push) edge node {$\mathit{Push_i}$} (hot1)
     (hot1) edge node {$\mathit{Hot_i}$} (hot2)
     (hot2) edge[bend right,midway,right] node {$\mathit{Hot_i}$} (hot3)
     (hot3) edge[bend right,midway,left] node {$\mathit{Hot_i}$} (push);
}

\item $\mathit{CBT}_\mathit{Int} \colon$ 
\cbt[my cbt style,
every state/.style={
    align=left,
    text width=20mm,
    rectangle,
    rounded corners,
},
node distance=4cm,
]
{Int}
{$Query=Kitchen$}
{
    \node[state,initial] (n1) [right=20pt of desc] {$R=\emptyset$\\$B=\{\mathit{Hot_i}\}$};
    \node[state,initial] (n2) [right of=n1] {$R=\emptyset$\\$B=\{\mathit{Cold_i}\}$};

    \path 
     (n1) edge [bend left, above] node {$\mathit{Cold_i}$} (n2)
     (n2) edge [bend left, below] node {$\mathit{Hot_i}$} (n1)
    ;
}
\end{itemize}

\subsection{Execution Demonstrations}
For demonstrating the execution of the above program, we denote $\mathit{LC}_{X,i}$ as a live copy of $\mathit{CBT}_X$ for room $i$, where $X \in \{\mathit{Cold},\mathit{Hot},\mathit{Int}\}$. We note that while $i$ is a parameter in the definition of the LTS of each CBT, when the live copies are spawned, their seed (i.e., a result to the CBT's query) contains the value of $i$.

In this example, we assume that there are three rooms: a kitchen, a bathroom, and a bedroom. We define our initial context to be 
$\{
 \langle 1, \mathit{kitchen} \rangle,
 \langle 2, \mathit{bathroom} \rangle,$ $\langle 3, bedroom \rangle
\}$, and our initial live copies are thus: $\{ \mathit{LC}_\mathit{Cold,1},  \mathit{LC}_\mathit{Hot,1},\allowbreak \mathit{LC}_\mathit{Int,1},\mathit{LC}_{Cold,2},\mathit{LC}_\mathit{Hot,2}\}$.
Each live copy uses a different subset of events, meaning that the behavior of each room with taps is independent with the behavior of other rooms. Therefore, we note two observations:
\begin{enumerate}
\item The interleave CBT is bound to the $\mathit{Kitchen}$ query. Thus, upon pushing the button of the kitchen and the bathroom, the kitchen's Hot/Cold events ($\mathit{Cold}_1$ and $\mathit{Hot}_1$) will be interleaved, and the bathroom's hot/cold events ($\mathit{Cold}_2$ and $\mathit{Hot}_2$) will not.
\item There are no constrains on the order of the events between the different rooms. Thus, for example, the following order is possible:
$\mathit{Push}_1,\mathit{Cold}_1,\mathit{Push}_2,\mathit{Hot}_2,\allowbreak \mathit{Hot}_1,\mathit{Hot}_2,\mathit{Cold}_1,\mathit{Hot}_1,\mathit{Cold}_2,\mathit{Cold}_1,\mathit{Hot}_2,\dots$
\end{enumerate}

\subsection{A Short Discussion}

\paragraph{Succinctness}
In some cases, COBP specifications are more succinct than BP specifications. In this example, we had three CBTs only, while a similar BP code will have two b-threads for each room with taps and an additional b-thread for each kitchen. \autoref{tab:succinctness} shows the correlation in BP between the number of rooms and the code length, as opposed to COBP, where the code length is unaffected.

\begin{table}[t]
\centering
\begin{tabular}{cc|c>{\bfseries}c}
\# of & \# of        & \# b-threads & \normalfont{\# CBTs} \\
bathrooms & kitchens & (BP) & \normalfont{(COBP)} \\\hline
0 & 0 & 0   & 3 \\
1 & 0 & 2   & 3 \\
0 & 1 & 3   & 3 \\
1 & 1 & 5   & 3 \\
2 & 1 & 7   & 3 \\
2 & 2 & 10  & 3 \\
10 & 10 & 50   & 3
\end{tabular}
\caption{The succinctness of a COBP specification for the extended hot-cold example, compared to a BP specification.}
\label{tab:succinctness}
\end{table}

\paragraph{Formal Verification}
Harel et al.~\cite{Harel2013ComposingProvingCorrectness} showed how verification of behavioral programs can be automated and streamlined by combining properties of individual modules, specified and verified separately. They demonstrated how their method may yield an exponential acceleration of the verification process when compared with model-checking the composite application.

In some cases, the context idioms generate individual modules that allow for using the compositional verification approach of~\cite{Harel2013ComposingProvingCorrectness} for verifying the behavior of a context-aware behavioral program. In this example, each answer to each of the two queries creates a set of live copies (two for a $\mathit{bathroom}$ and three for a $\mathit{kitchen}$). According to the implementation, each set uses its own set of events (on the edges, in $R$, or in $B$) --- $\{\mathit{Push_i},\mathit{Hot}_i,\mathit{Cold}_i \colon i \in \mathbb{N}\}$. In addition, the effect function does not change the context. Thus, each set is an independent module that does not affect the behavior of other sets. 

Based on these observations, to verify the consistency of the program behavior with the kitchens' interleave requirement (requirement 4), it is sufficient to use one kitchen only. For that, we will show that for a single kitchen, the LTS of \autoref{def:context:run} is equivalent to the LTS of \autoref{def:bt-run}. Since it is possible to verify behavioral programs, it is also possible to verify this program. Given the initial state $\langle \{\langle 1, \mathit{kitchen} \rangle\},\{ \mathit{LC}_\mathit{Cold,1},  \mathit{LC}_\mathit{Hot,1}, \mathit{LC}_\mathit{Int,1}\}\rangle$, \autoref{def:context:run} is equivalent to \autoref{def:bt-run}, since: (1) \autoref{def:context:run}[3.a] $\equiv$ \autoref{def:bt-run}[1]; (2) $\mathit{ctx}' =\allowbreak \mathit{effect}(e)(\mathit{ctx})=\mathit{ctx}$; (3) $\mathit{LC}'=  \mathit{LC}'_\mathit{run}$; and (4) $\mathit{LC'}_{run}\equiv$ \autoref{def:bt-run}[2]. 

This simple example shows how a COBP program can be verified using the abstract semantics. In addition it shows how the context idioms increase the modularity of the program, thus allow for exponential acceleration of the verification process. In \autoref{sec:case-studies:ros:disscussion} we demonstrate the verification process of COPB program written in JavaScript, and in \autoref{sec:related-work} we survey additional reasoning techniques for scenario-based programming.

\section{Example --- The Game of Life}
\label{sec:examples-life}
The Game of Life, or simply Life, is a zero-player game based on a cellular automaton, devised by Conway in 1970~\cite{Gardner1970MathematicalGames}. The game board is an infinite, two-dimensional orthogonal grid of cells, where each cell can be either populated with an individual or unpopulated. Each individual interacts only with its neighbors, that is, individuals that are in one of the eight cells adjacent to it. The user sets the initial population of the game. Then, the following four rules are applied to the individuals of the first generation to generate the population of the first generation: 
\begin{enumerate}
\item Any individual with fewer than two neighbors --- dies --- as if by underpopulation.
\item Any individual with two or three neighbors --- lives on to the next generation.
\item Any individual with more than three neighbors --- dies --- as if by overpopulation.
\item Any three individuals that are the only ones that surrounds an unpopulated cell --- reproduce an individual at that cell.
\end{enumerate}

All rules are applied simultaneously to all of the individuals --- births and deaths occur simultaneously, at a discrete moment called a tick. The rules are repeatedly applied to create further generations. 

We note that each requirement depends on a different aspect of the context: for the first requirement --- ``individual with fewer than two neighbours'', for the second requirement --- ``individual with two or three neighbours'', and so on. 

\subsection{Implementation}
\label{sec:examples-life:impl}
We define the possible events as $E=\{\mathit{die(row,col)},\allowbreak \mathit{reproduce(row,col)},\mathit{tick},\mathit{tock} \colon \mathit{row},\mathit{col} \in \mathbb{N}\}$, where the ev\-ents $\mathit{die}(\mathit{row},\mathit{col})$ and $\mathit{reproduce(row,col)}$ represent the death and reproduce (respectively) of an individual at the cell $\langle \mathit{row,col} \rangle$. Finally, to synchronize the creation of live copies we used $\mathit{tick}$ to represent the state of the system at the beginning moment of a generation, and $\mathit{tock}$ to represent the rest of the system states.

\subsubsection*{Context Specification}
\label{sec:examples-life:context-spec}
\begin{itemize}[itemsep=1ex]
\item We define the set of all possible system contexts, $\mathit{CTX}$, to be $\{ \langle \mathit{Pop,Tick} \rangle \colon \mathit{Pop} \subseteq \{\langle \mathit{row,col} \rangle \colon \mathit{row,col}\allowbreak \in \mathbb{N}\}, \mathit{Tick} \in \{0,1\}\}$, where the population (represented by $\mathit{Pop}$) is a set of locations of all living individuals, and $\mathit{Tick}=1$ if and only if all of the individuals have finished processing their rules for the current generation, and the rules-processing of the following generation has not yet started.

\item The set of all cells that can be reached from the cell $\langle \mathit{row,col} \rangle$ in one step, is defined by the function $\mathit{ngb(row,col)} = \{\langle \mathit{row}+1,\mathit{col} \rangle, \langle \mathit{row}+1,\mathit{col}+1 \rangle, \allowbreak \langle \mathit{row},\mathit{col}+1 \rangle,\langle \mathit{row}-1,\mathit{col}+1 \rangle,\langle \mathit{row}-1,\mathit{col} \rangle, \allowbreak \langle \mathit{row-1, \allowbreak col-1} \rangle,\allowbreak \langle \mathit{row},\mathit{col}-1 \rangle,\langle \mathit{row}+1,\mathit{col}-1 \rangle\}$. 
\end{itemize}

\subsubsection*{Data-Access Layer}
\label{sec:examples-life:dal1}
As noted before, each requirement is bound to a different context. Therefore, for each context we define a different query, labeling it by the requirement ID:

\begin{itemize}[itemsep=1ex, label={}, leftmargin=0pt]
\item \scaletoline{$Q_1=\tickPartedQuery[
\left\{\begin{array}{lcl}
    \multicolumn{3}{l}{\langle row,col \rangle \colon} \\ 
    & \multicolumn{2}{l}{\langle row,col \rangle \in Pop} \\
    & \land & \lvert ngb(row,col) \cap Pop \rvert < 2
\end{array}\right\}]$}

\item \scaletoline{$Q_2=\tickPartedQuery[
\left\{\begin{array}{lcl}
    \multicolumn{3}{l}{\langle row,col \rangle \colon} \\ 
    & \multicolumn{2}{l}{\langle row,col \rangle \in Pop} \\
    & \land & 2 \leq \lvert ngb(row,col) \cap Pop \rvert \leq 3
\end{array}\right\}]$}

\item \scaletoline{$Q_3=\tickPartedQuery[
\left\{\begin{array}{lcl}
    \multicolumn{3}{l}{\langle row,col \rangle \colon} \\ 
    & \multicolumn{2}{l}{\langle row,col \rangle \in Pop} \\
    & \land & \lvert ngb(row,col) \cap Pop \rvert > 3
\end{array}\right\}]$}

\item \scaletoline{$Q_4=\tickPartedQuery[
\left\{\begin{array}{lcl}
    \multicolumn{3}{l}{\langle row,col \rangle \colon} \\ 
    & \multicolumn{2}{l}{\langle row,col \rangle \notin Pop} \\
    & \land & \lvert ngb(row,col) \cap Pop \rvert = 3
\end{array}\right\}]$}
\end{itemize}

The effect function is:
\begin{itemize}[itemsep=1pt]
\item $\mathit{die(row,col)}(\langle \mathit{Pop,Tick} \rangle) = \langle \mathit{Pop} \setminus \{\langle \mathit{row,col} \rangle \},\allowbreak \mathit{Tick}\rangle$, is the effect of the event $die(row,col)$ that removes the individual from the population.

\item $\mathit{reproduce(row,col)}(\langle \mathit{Pop,Tick} \rangle)=\langle \mathit{Pop} \cup \{\langle \mathit{row,col}\allowbreak \rangle \}, Tick\rangle$, is the effect of the event $reproduce(row,col)$, adding the reproduced individual to the population.

\item $\mathit{tick}(\langle \mathit{Pop,Tick} \rangle) = \mathit{tock}(\langle \mathit{Pop,Tick} \rangle) = \langle \mathit{Pop},1-\mathit{Tick} \rangle$, are the effects of the events $\mathit{tick}$ and $\mathit{tock}$.
\end{itemize}

\subsubsection*{Behavioral Specification (CBTs)}
\label{sec:examples-life:cbts}
For each rule, we define a context-aware b-thread, called $\mathit{CBT}_i$ as the $\mathit{CBT}$ for the i-th rule, where $Q_i$ is the definition of its query, and $\mathit{BT}_i$ is the definition for its tuple  $\langle S,\rightarrow, \mathit{init}, R, B \rangle$.

\begin{itemize}[leftmargin=0pt, label={}]
\item $\mathit{CBT}_1\colon$ 
\golcbt{1}{Q_1}{
    \node[state,initial] (n1) [below=0mm of desc] {$R=\mathit{die(row,col)}$\\$B=\{\mathit{tick}\}$};
    \node[state,text width=13mm] (n2) [right of=n1] {$R=\emptyset$\\$B=\emptyset$};
    \path 
     (n1) edge[loop below] node{otherwise} (n1)
     (n1) edge node{$\mathit{die(row,col)}$} (n2);
}

\item $\mathit{CBT}_2\colon$ 
\golcbt{2}{Q_2}{
    \node[state,initial,text width=13mm] (n1) [below=0cm of desc] {$R=\emptyset$\\$B=\emptyset$};
}

\item $\mathit{CBT}_3\colon$ 
\golcbt{3}{Q_3}{
    \node[state,initial] (n1) [below=0cm of desc] {$R=\mathit{die(row,col)}$\\$B=\{\mathit{tick}\}$};
    \node[state,text width=13mm] (n2) [right of=n1] {$R=\emptyset$\\$B=\emptyset$};
    \path 
     (n1) edge[loop below] node{otherwise} (n1)
     (n1) edge node{$\mathit{die(row,col)}$} (n2);
}

\item $\mathit{CBT}_4\colon$ 
\golcbt{4}{Q_4}{
    \node[state,initial,text width=38mm] (n1) [below=0mm of desc] {$R=\mathit{reproduce(row,col)}$\\$B=\{\mathit{tick}\}$};
    \node[state,text width=10mm] (n2) [below=1cm of n1] {$R=\emptyset$\\$B=\emptyset$};
    \path 
     (n1) edge[loop above] node{otherwise} (n1)
     (n1) edge node{$\mathit{reproduce(row,col)}$} (n2);
}
\end{itemize}

\vspace{1em}\noindent To synchronize the generation tick, we define:

\begin{itemize}[leftmargin=0pt, label={}]
\item $\mathit{CBT}_{tick} \colon$ 
\golcbt{tick}{\{1\}}{
    \node[state,initial] (n1) [below=0cm of desc] {$R=\{tick\}$\\$B=\emptyset$};
    \node[state] (n2) [right=5mm of n1] {$R=\{tock\}$\\$B=E \setminus \{tock\}$};
    \path 
        (n1) edge[loop below] node{otherwise} (n1)
        (n2) edge[loop below] node{otherwise} (n2)
        (n1) edge [bend left, above] node {$\mathit{tick}$} (n2)
        (n2) edge [bend left, below] node {$\mathit{tock}$} (n1)
    ;
}
\end{itemize}

\subsection{Execution Demonstrations}
We now demonstrate the execution of the program with different initial populations. We do that by presenting how the context and the live copies change over the generations. For each generation $i$, we present the set of the live copies at their initial state, denoted by $\mathbb{LC}_i$, and the state of the context at the beginning of the generation, denoted by $ctx_i$. Given $name \in \{1,2,3,4,tick\}$, we denote $\mathit{LC}_{name,\langle row,col \rangle}=\langle \mathit{CBT}_{name},Q_{name},\langle row,col \rangle \rangle$ as the live copy of $\mathit{CBT}_{name}$ where the result of $Q_{name}$ (i.e., the seed --- $c$), is $\langle row,col \rangle$. 

\paragraph{Two Lonely Individuals}\mbox{}
In this example, the seed generation has two individuals that are not neighbors. According to the rules, they must die in the following iteration, i.e., the answer for $Q_1$ contains the two individual (and only them):

$ctx_0 = \langle \{\langle 5,5 \rangle\, \langle 10,10 \rangle\}, 0 \rangle$

$\mathbb{LC}_0 = \{ \mathit{CBT}_{tick}, LC_{1,\langle 5,5 \rangle}, LC_{1,\langle 10,10 \rangle} \}$

$\mathbb{LC}_1 = \emptyset$

\paragraph{Three Individuals In a Row (Blinker)}\mbox{}
There are interesting patterns in Life, categorized by the behavior of the individuals. The \emph{oscillators} category includes patterns that the individuals return to their initial state after a finite number of generations. One of these patterns is the \emph{Blinker}, where three individuals are ordered in a horizontal line, with no empty cells between them. On even generations the individuals return to their initial state, and on odd generations the individuals change to a vertical position. Thus, the execution will be:

\noindent $ctx_{2i} = \langle \{\langle 5,4 \rangle\, \langle 5,5 \rangle\, \langle 5,6 \rangle\}, 0 \rangle$

\noindent$\begin{array}{@{}l@{}l}
\mathbb{LC}_{2i} = \{ & \mathit{CBT}_{tick}, LC_{1,\langle 5,4 \rangle}, LC_{2,\langle 5,5 \rangle}, LC_{1,\langle 5,6 \rangle},\\
& LC_{4,\langle 4,5 \rangle}, LC_{4,\langle 6,5 \rangle} \}
\end{array}$

\noindent $ctx_{2i+1} = \langle \{\langle 4,5 \rangle\, \langle 5,5 \rangle\, \langle 6,5 \rangle\}, 0 \rangle$

\noindent$\begin{array}{@{}l@{}l}
\mathbb{LC}_{2i+1} = \{ & \mathit{CBT}_{tick}, LC_{1,\langle 4,5 \rangle}, LC_{2,\langle 5,5 \rangle}, LC_{1,\langle 6,5 \rangle},\\
& LC_{4,\langle 5,4 \rangle},\allowbreak LC_{4,\langle 5,6 \rangle} \}
\end{array}$

\section{Example --- Evolved Game of Life}
While the CBTs for the game rules demonstrate the concept of context, they consist of a single state, thus missing the `scenario' trait of BP. 

\begin{figure}[ht]
  \centering
  \includegraphics[width=0.15\textwidth]{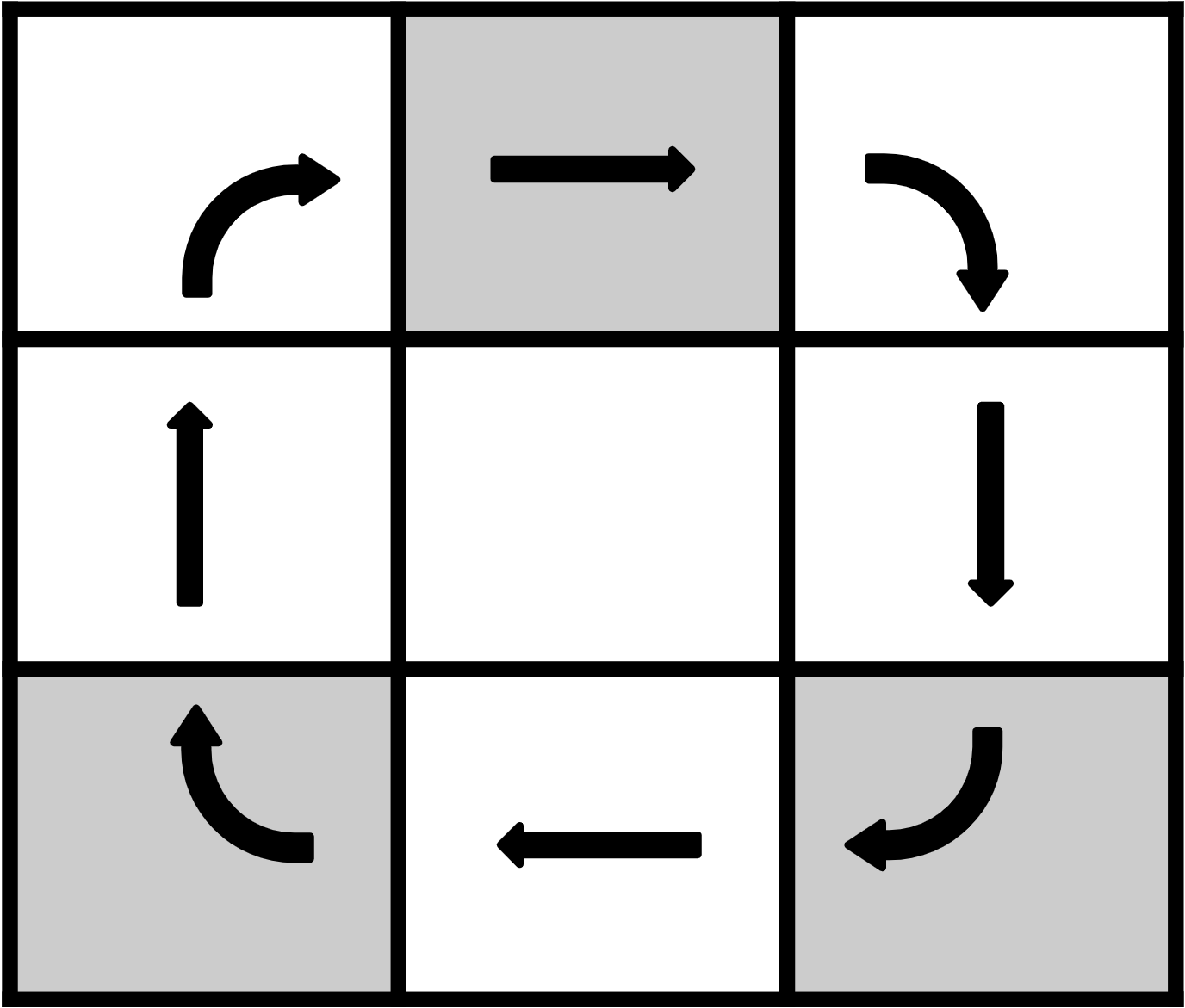}
  \caption{A mating dance. The three, lonely gray individuals surround the unpopulated cell at the center. At each of the following eight generations, the three individuals will move one step along the arrows, from one cell to another, until returning to their original position, completing the mating dance. Then, they will reproduce an individual at the center cell.}
  \label{fig:mating_dance}
\end{figure}

Consider for example the following new requirements for the game (in analogy to the \emph{Laws of Robotics} by Asimov~\cite{Asimov1942Runaround}):

\begin{enumerate}[label=\Alph*.]
\item Any three \textbf{lonely} individuals (i.e., individuals with no neighbors) that are the only ones that surround an unpopulated cell --- perform the \emph{mating dance} --- a complete clockwise circling around the unpopulated cell (depicted in~\autoref{fig:mating_dance}). Upon completion, the three reproduce an individual at that cell.
\item \emph{otherwise}, the original rules are valid:
\begin{enumerate}[label=\arabic*.]
\item Any individual with fewer than two neighbors --- dies --- as if by underpopulation.
\item Any individual with two or three neighbors --- lives on to the next generation.
\item Any individual with more than three neighbors --- dies --- as if by overpopulation.
\item Any three individuals that are the only ones that surrounds an unpopulated cell --- reproduce an individual at that cell.
\end{enumerate}
\end{enumerate}

We note that rule A defines a new context-dependent requirement, while B rules only refine the context of the original rules.

\subsection{Implementation}
The mating dance involves eight `dancing' steps of three lonely individuals around an unpopulated cell $\langle row,col \rangle$, until they complete a circle. We represent the dancing step of the three individuals with the event $step(row,col)$, that is added to the previously defined event-set (in~\autoref{sec:examples-life:impl}).

One way to implement the B requirements, is to add a condition to each of the CBTs of the original system, that validates that the individual does not participate in a mating dance. We will take a more subtle approach that is made possible by the COBP paradigm. Specifically, since only the context of the original requirements has changed, and the behavior did not, we use the original CBTs and only change their queries to exclude the case of a mating dance. 

\subsubsection*{Context Specification}
No changes.

\subsubsection*{Data-Access Layer}
\label{sec:examples-life:dal2}
To simplify $Q_A$, we define the function $\mathit{isLonely}$ that receives a cell and returns $\mathit{true}$ iff it has no neighbors:\\
$isLonely(row,col) = true \Longleftrightarrow (\lvert ngb(row,col) \cap Pop \rvert = 0)$.

Our queries are then:
\begin{itemize}[leftmargin=*]
\setlength{\itemsep}{1.5\baselineskip}
\item {\scaletoline{ $Q_A=\tickPartedQuery[
\left\{\begin{array}{lll@{\hspace{8pt}}l}
    \multicolumn{4}{l}{\langle row,col \rangle \colon} \\ 
    & \multicolumn{3}{l}{\langle row,col \rangle \notin Pop}\\
    & \land & \multicolumn{2}{l}{\lvert ngb(row,col) \cap Pop \rvert = 3} \\
    & \land & \multicolumn{2}{l}{\forall \langle i,j \rangle \in ngb(row,col) \cap Pop \colon} \\
    & & & isLonely(i,j)
\end{array}\right\}]$}}
\item For each of the original queries in~\autoref{sec:examples-life:dal1}, we add the following constraint to make sure that the individuals do not participate in a mating dance: $\nexists \langle i,j \rangle \in Q_A \colon \langle row,col \rangle \in ngb(i,j)$. The new queries are:

\begin{itemize}[label=,leftmargin=*]
\item  \scaletoline{$Q_{B1}=\tickPartedQuery[
\left\{\begin{array}{lcl}
    \multicolumn{3}{l}{\langle row,col \rangle \colon} \\ 
    & \multicolumn{2}{l}{\nexists \langle i,j \rangle \in Q_A \colon \langle row,col \rangle \in ngb(i,j)} \\
    & \land & \langle row,col \rangle \in Pop \\
    & \land & \lvert ngb(row,col) \cap Pop \rvert < 2
\end{array}\right\}]$}

\item \scaletoline{$Q_{B2}=\tickPartedQuery[
\left\{\begin{array}{lcl}
    \multicolumn{3}{l}{\langle row,col \rangle \colon} \\ 
    & \multicolumn{2}{l}{\nexists \langle i,j \rangle \in Q_A \colon \langle row,col \rangle \in ngb(i,j)} \\
    & \land & \langle row,col \rangle \in Pop \\
    & \land & 2 \leq \lvert ngb(row,col) \cap Pop \rvert \leq 3
\end{array}\right\}]$}

\item \scaletoline{$Q_{B3}=\tickPartedQuery[
\left\{\begin{array}{lcl}
    \multicolumn{3}{l}{\langle row,col \rangle \colon} \\ 
    & \multicolumn{2}{l}{\nexists \langle i,j \rangle \in Q_A \colon \langle row,col \rangle \in ngb(i,j)} \\
    & \land & \langle row,col \rangle \in Pop \\
    & \land & \lvert ngb(row,col) \cap Pop \rvert > 3
\end{array}\right\}]$}

\item \scaletoline{$Q_{B4}=\tickPartedQuery[
\left\{\begin{array}{lcl}
    \multicolumn{3}{l}{\langle row,col \rangle \colon} \\ 
    & \multicolumn{2}{l}{\nexists \langle i,j \rangle \in Q_A \colon \langle row,col \rangle \in ngb(i,j)} \\
    & \land & \langle row,col \rangle \notin Pop \\
    & \land & \lvert ngb(row,col) \cap Pop \rvert = 3
\end{array}\right\}]$}
\end{itemize}
\end{itemize}

To the the effect function we add an effect for the $\mathit{step(row,col)}$ event, that advances the neighbors to their next dance location: \vspace{1em}

\noindent\scaletoline{{$\begin{array}{ll}
\multicolumn{2}{l}{step(row,col)(\langle Pop, Tick \rangle)=} \\[1em]
& \langle \underbrace{(Pop \setminus ngb(row,col))}_{\text{Remove old locations}} \cup \\[2em]
& \underbrace{\{next(row,col,i,j) \colon \langle i, j \rangle \in ngb(row,col)) \cap Pop \}}_{\text{Add next locations}}, Tick \rangle
\end{array}$}}\vspace{1em}

\noindent where $next(row,col,i,j)$ is a function that returns the consecutive cell to $\langle i,j \rangle $ in the mating dance circle. For example, $next(row,col,row,col+1)=\langle row,col,row+1,col+1 \rangle$, $next(row,col,row+1,col+1)=\langle row,col,row+1,col \rangle$, and so forth.

\subsubsection*{Behavioral Specification (CBTs)}
Since the behavior of the original CBTs did not change, only their context, we only change the query names of the CBTs (i.e., $Q_1$ will now be $Q_{B1}$, $Q_2$ will now be $Q_{B2}$, etc.). We define the following additional CBT for rule A:

{\scriptsize
\begin{itemize}[leftmargin=0pt]
\item[] $\mathit{CBT}_A$: 
\golcbt{A}{}{
    \node[state,initial] (n1)[text width=26mm, below=0cm of desc] {$R=\{\mathit{step(row,col)}\}$\\$B=\{\mathit{tick}\}$};
    \node[state] (n2) [text width=26mm,right=2.1cm of n1] {$R=\{\mathit{step(row,col)}\}$\\$B=\{\mathit{tick}\}$};
    \node[state] (n9) [text width=26mm,below=1cm of n1] {$R=\{\mathit{step(row,col)}\}$\\$B=\{\mathit{tick}\}$};
    \node[state] (quit) [text width=10mm,below=1cm of n2] {$R=\emptyset$\\$B=\emptyset$};
    
    \path 
     (n1) edge[loop above] node{otherwise} (n1)
     (n2) edge[loop above] node{otherwise} (n2)
     (n9) edge[loop below] node{otherwise} (n9)
     
     (n1) edge node {$\mathit{step(row,col)}$} (n2)
     (n9) edge node{$\mathit{reproduce(row,col)}$} (quit)
     ;
     \draw [->] (n2.south)  --([shift={(0cm,-0.55cm)}]n2.south)-- ([shift={(0cm,0.45cm)}]n9.north) node [midway,above=-2ex,fill=white] {$\cdots \mathit{step(row,col)} \times 7 \cdots $} -|  (n9);
}
\end{itemize}
}

\subsection{Execution Demonstrations}

\begin{figure}[t]
    \centering
    \quad\quad
    \subfloat[Generation 0]{
        \includegraphics[width=0.23\linewidth]{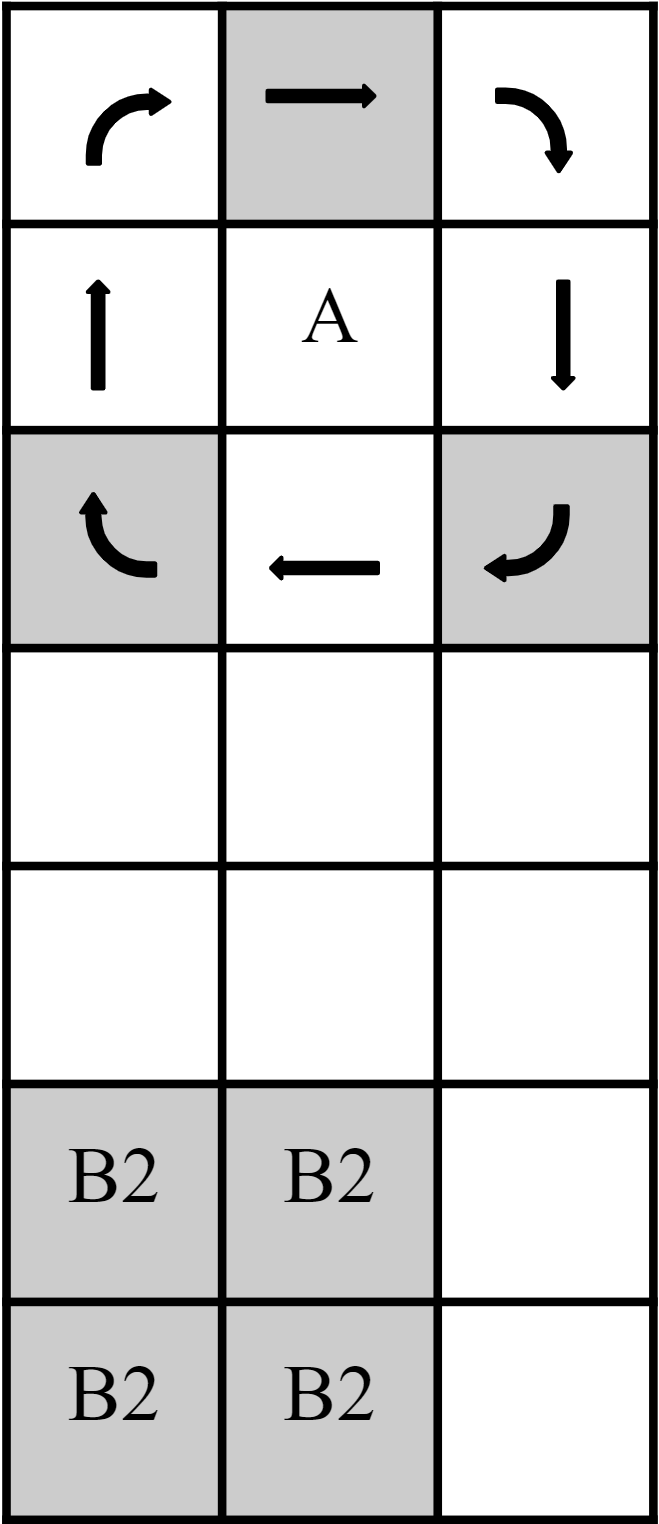}
    }
    \hfill
    \subfloat[Generation 1]{
        \includegraphics[width=0.23\linewidth]{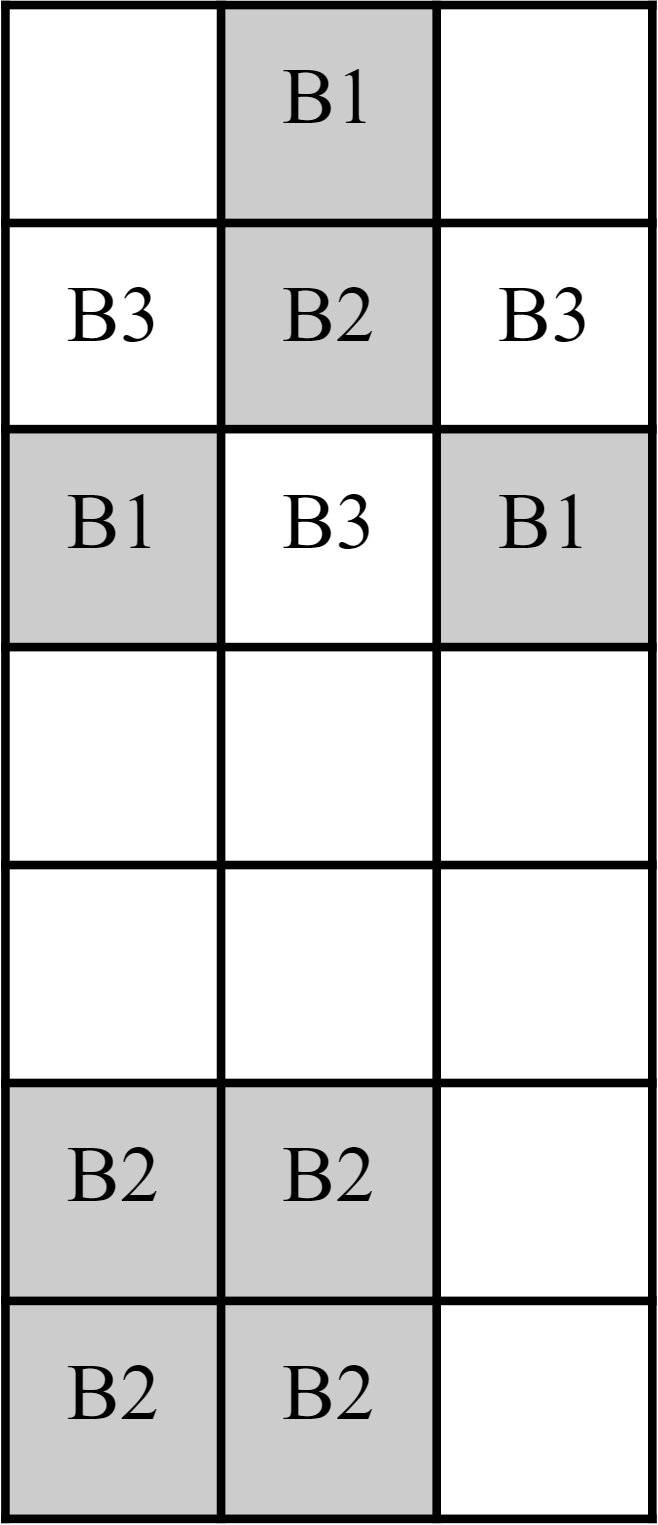}
    }
    \hfill
    \subfloat[Generation 2]{
        \includegraphics[width=0.23\linewidth]{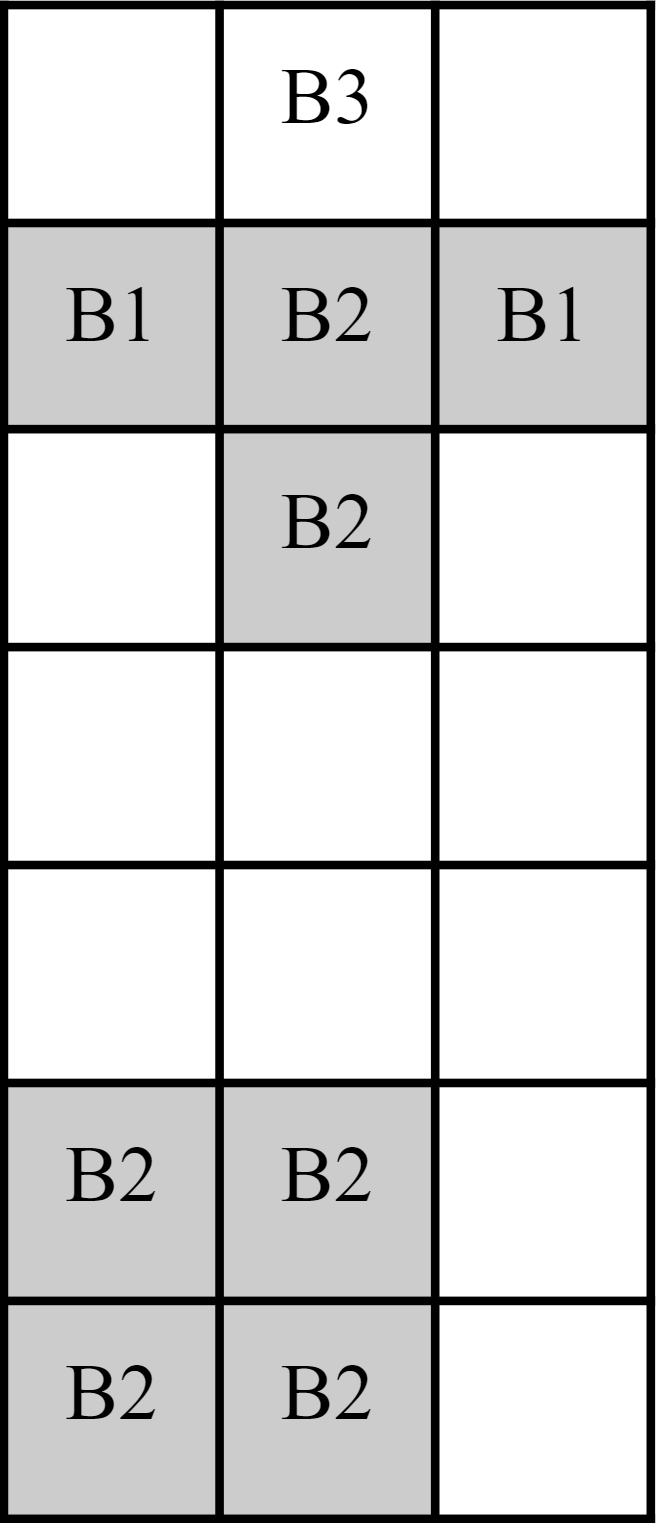}
    }
    \quad\quad
    \caption{The individuals (in gray) at the beginning of the first three generation of the extended Game of Life example. For each cell/individual that one of the rules is valid for --- the rule number is stated. The block pattern is at the bottom of the grid, and the mating-dance pattern is at the top of the grid at generation 0. The dancing takes place during generation 0, and at the end of the dance, a new individual is spawned at cell $\langle 1,1\rangle$.}
    \label{fig:mating_dance-run}
\end{figure}

\paragraph{Mating dance and still life}\mbox{}
In this example, depicted in~\autoref{fig:mating_dance-run}, we have two patterns. At the bottom of the grid, is the \emph{block} pattern of the \emph{still life} type. It is called `still life' since it does not change over generations. The second pattern is our new \emph{dancing} pattern. We demonstrate the run for three generations, presenting only the initial state of each generation. The dance starts and completes during the first generation. 

Since the individuals of the block pattern do not change over generations, at each generation a new live copy of $\mathit{CBT}_{B2}$ is generated for each one of them. Thus, for simplification, we denote the individuals as $\mathit{block}=\{\langle 5,0 \rangle, \langle 5,1 \rangle,\allowbreak \langle 6,0 \rangle, \langle 6,1 \rangle \}$, and their live copies as $\mathbb{LC}_\mathit{block} = \{ \mathit{LC}_{B2,\langle 5,0 \rangle},\allowbreak \mathit{LC}_{B2,\langle 5,1 \rangle}, \mathit{LC}_{B2,\langle 6,0 \rangle}, \mathit{LC}_{B2,\langle 6,1 \rangle}\}$.

\begin{description}[itemsep=1ex,leftmargin=1em]
\item Generation 0:

$\mathit{ctx}_0 = \langle \{\langle 0,1 \rangle, \langle 2,0 \rangle, \langle 2,2 \rangle \} \cup \mathit{block}, 0 \rangle$
    
$\mathbb{LC}_0 = \{ \mathit{CBT}_\mathit{tick}, \mathit{LC}_{A,\langle 1,1 \rangle} \} \cup \mathbb{LC}_\mathit{block}$
    
\item Generation 1:

$\mathit{ctx}_1 = \langle \{ \langle 0,1 \rangle, \langle 1,1 \rangle, \langle 2,0 \rangle, \langle 2,2 \rangle \} \cup \mathit{block}, 0 \rangle$

\scaletoline{$\begin{array}{@{}c@{}l}
     \mathbb{L}&\mathbb{C}_1 = \{ \mathit{CBT}_\mathit{tick}, \mathit{LC}_{B1,\langle 0,1 \rangle}, \mathit{LC}_{B3,\langle 1,0 \rangle}, LC_{B2,\langle 1,1 \rangle}, \\
     & \mathit{LC}_{B3,\langle 1,2 \rangle}, \mathit{LC}_{B1,\langle 2,0 \rangle}, \mathit{LC}_{B3,\langle 2,1 \rangle}, \mathit{LC}_{B1,\langle 2,2 \rangle} \} \cup \mathbb{LC}_\mathit{block}
\end{array}$}

\item Generation 2:

$ctx_2 = \langle \{\langle 1,0 \rangle, \langle 1,1 \rangle, \langle 1,2 \rangle, \langle 2,1 \rangle \} \cup block, 0 \rangle$
    
$\begin{array}{@{}c@{}l}
     \mathbb{L}&\mathbb{C}_2 = \{ \mathit{CBT}_{tick}, LC_{B3,\langle 0,1 \rangle}, LC_{B1,\langle 1,0 \rangle}, LC_{B2,\langle 1,1 \rangle}, \\
     & LC_{B1,\langle 1,2 \rangle}, LC_{B2,\langle 2,1 \rangle} \cup \mathbb{LC}_{block}
\end{array}$

\end{description}

\subsection{A Short Discussion}
Our initial definition for requirement A did not require that the three individuals will be lonely. When we implemented and run the last example, we observed a bug --- in some cases, during the mating dance, the individuals have been duplicated and each copy stepped into a different cell. We added an assertion that checks this run property, and because of the formal semantics of the COBP paradigm, we were able to verify the existence of the bug. Moreover, we were able to find an initial seed that causes the bug --- when the three cells are ordered in a vertical or a horizontal line, then they are neighbors of two unpopulated cells (i.e., from the two sides of the line). In fact, we discovered that the bug was in the requirements which is why we changed them to three lonely individuals. We provide a detailed description of the verification process and related experiments that we have conducted in~\autoref{sec:case-studies:iot:discussion}. 

\section{Case Studies --- Motivation and Overview}
\label{sec:case-studies:motivation}
We now turn to present two case studies of two different industrial fields --- robotics and the internet-of-things. In both case studies we developed the systems using our BPjs-based implementation, described in~\autoref{sec:cobp:implementation}.

The purposes of these case studies are: to present an implementation of the paradigm and a syntax for the abstract semantics; to discuss additional implementation details and design patterns; and to demonstrate real-life use cases where the paradigm may excel other paradigms, both context oriented and not. To keep the focus of this paper, we only discuss the key differences between our implementation and possible implementations in other paradigms and languages. Additional aspects are discussed in~\autoref{sec:related-work}.

The contextual data is stored in a relational database (an in-memory SQLite database\footnote{SQlite –\url{https://www.sqlite.org}}) and managed by a data-access layer (DAL) using the Hibernate ORM framework\footnote{Hibernate ORM --- \url{https://hibernate.org/orm}}. In analogy to ``database views'', the context's `select' queries allow for triggering a stored procedure whenever a record is added or removed from its result (i.e., announcing the changes via events and spawning live copies upon new records) . The context's `update' commands update the data as required (by adding, deleting, and changing objects and object relations). Query and update commands are accessed by their IDs in the business-logic layer (i.e., the behavioral specification defined by the CBTs), leaving the context-related implementation decisions to the lower layer (i.e., the DAL and the DB). 

This implementation can be found at~\url{https://github.com/bThink-BGU/BPjs-Context} and the following case studies can be found at~\url{https://github.com/bThink-BGU/Papers-2020-COBP}. 

\section{Case Study --- Robotic Operating System (ROS)}
\label{sec:case-studies:ros}
Modern autonomous robots, such as smart manipulators and self-driving cars and drones, run many parallel tasks such as obstacle avoidance, exploration, mapping, and navigation. Concurrently, they also need to plan ahead, do fault detection, and preserve their integrity. Many service applications such as, e.g., home-assistance robots, surveillance robotics, inspection, rescue robotics, or entertainment robotics require to achieve several goals at the same time. The goals may conflict with one another, and the significance of a task is frequently context dependent. For instance, in a car, avoiding a far obstruction may be of little significance relative to achieving a nearby target position.


\begin{figure}[t]
    \centering
    \includegraphics[width=0.7\linewidth]{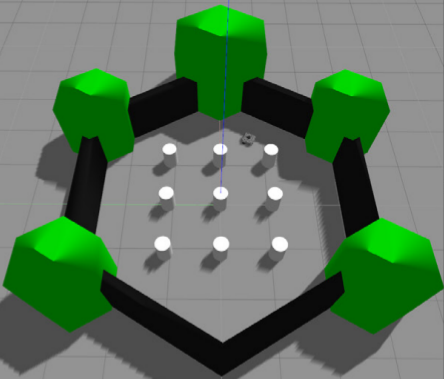}
    \caption{The TurtleBot3 world (credit:~\url{http://emanual.robotis.com}). The Robot must not hit the walls while moving around.}
    \label{fig:turtlebot}
\end{figure}

In this case study, we demonstrate how COBP can be used for developing decision-making components in ROS-based robotic systems. The Robot Operating System (ROS) framework aims to simplify the task of creating complex and robust robot behavior. At the lowest level, ROS offers a message passing interface that provides an inter-process communication. On top of it, common robot-specific libraries and tools are supplied to get the robot up and running quickly. While ROS simplifies the task of developing robots, it does not include built-in tools for developing the complex and robust robot behaviors. Thus, decision-making components are developed in various ways, such as coded using imperative languages (e.g., C++ or Python); modeled using third-party libraries, such as behavior-tree libraries (e.g., \url{http://wiki.ros.org/decision_making}); or, learned by means of artificial intelligence~\cite{Kaminka2018PlanRecognitionContinuous}. Behavior tree is a plan execution, describing the system behavior by switching between a finite set of tasks~\cite{Colledanchise2017BehaviorTreesRobotics}. They are popular among robots developers and they are considered as a ``very efficient way of creating complex systems that are both modular and reactive''~\cite{Colledanchise2017BehaviorTreesRobotics}. While that may be true, some behaviors are defined more easily by describing what must not happen. Consider for example a robot, designed to do tasks while avoiding obstacles, or ignoring certain actions when the battery is low. Adding such conditions may cause the tree to grow unwieldy. Refactoring conditions as hierarchy nodes, or composing multiple trees, address the size problem, but may cause additional problems as complex nodes and undesired behaviors. 

To better capture the idea of how COBP can be used for developing robots, we have implemented a COBP version of the decision-making component, taken from the official ROS tutorial\footnote{The official ROS tutorial~\url{http://wiki.ros.org/turtlebot3_simulations}}. The component defines how to move the robot around, while avoiding collision with the walls (see~\autoref{fig:turtlebot}). Part of the tutorial code for this decision-making component is presented in~\autoref{lst:ros-cpp}, written in C++. This \emph{specific} implementation raises many questions: Is it readable? Is it understandable? Can we easily adjust the code upon new requirements (e.g., do not turn right more than three times in a row)? Can we reuse this code for different tasks (e.g., if we wish to both avoid walls and reach four coordinates in a certain order)? While there are better practices for implementing this robot, a context-oriented approach has several advantages such as: a higher level of modularity of the behavioral specification (i.e., behaviors are bound to contexts); a better alignment to the requirements, a better agility (i.e., changing the context of a requirement only changes the context of the b-thread, a separation of concerns between the behavior and the data structure, etc.); and more.

\begin{lstlisting}[
  language=C++,
  label={lst:ros-cpp},
  caption={Part of the tutorial code for the decision-making component of the TurtleBot simulation.},
  float=t,
  breaklines=false, 
%   xleftmargin=.21\textwidth,
%   xrightmargin=.21\textwidth,
]
bool Turtlebot3Drive::controlLoop() {
  switch(state) {
    case GET_TB3_DIRECTION:
      if (distToObstacle[CENTER] > min_forward_dist) {
        if (distToObstacle[LEFT] < min_side_dist) {
          previous_position = position;
          state = TB3_RIGHT_TURN;
        } else if (distToObstacle[RIGHT] < min_side_dist){
          previous_position = position;
          state = TB3_LEFT_TURN;
        } else { state = TB3_DRIVE_FORWARD; }
      }
      if (distToObstacle[CENTER] < min_forward_dist) {
        previous_position = position;
        state = TB3_RIGHT_TURN;
      } break;
    case TB3_DRIVE_FORWARD:
      Move(LINEAR_VELOCITY, 0.0);
      state = GET_TB3_DIRECTION;
      break;
    case TB3_RIGHT_TURN:
    ...
\end{lstlisting}

\subsection{Implementation}
The possible events are $\{\mathit{move}(l,a), \allowbreak scan(\langle d_1, \cdots ,\allowbreak d_{360} \rangle),\allowbreak \mathit{CTX.Ended}(q,c)\}$, where $\mathit{move}$ represents moving the robot forward with a linear velocity of $l$ and rotating it in with an angular velocity of $a$, $\mathit{scan}$ represents the distances from the robots to obstacles in 360 degrees, and $\mathit{CTX.Ended}$ declares that $c$ is no longer the answer to query $q$. 

\subsubsection*{Context Specification}
Since this problem is extremely simple, the context schema (that defines $\mathit{CTX}$) is specified by a single table, called $\mathit{robot}$, with four attributes --- an id (primary key), and three numbers, called $\mathit{oAhead}$, $\mathit{oLeft}$, and $\mathit{oRight}$, that represent the distance from the robot to an obstacle in that direction (if there is no obstacle, the value is infinite). Since there is only one robot, the table will contain only one row, though it supports more.

\subsubsection*{Data Access Layer}
$\mathit{QUERY}$ and $\mathit{UPDATE}$ are defined in~\autoref{tab:case-studies:ros}. The effect function of the $\mathit{scan}$ event triggers a call to the update function $\mathit{SetObstacles}$ with the relevant data ---\\ \lstinline[prebreak=]|scan($\langle d_1, \cdots , d_{360} \rangle$)(ctx) = SetObstacles({ "a":$d_\mathit{CENTER}$,"l":|\\
\lstinline[prebreak=]|$d_\mathit{LEFT}$, "r":$d_\mathit{RIGHT}$ })|.

\begin{table}
    \centering
\begin{tabularx}{\linewidth}{@{}c@{\hspace{1ex}}l@{\hspace{1ex}}X@{}}
    \textbf{Type}   & \textbf{Name}   & \textbf{Command}   \\\hline
    Q   & $\mathit{Robot}$   & \lstinline[language=SQL]|SELECT * FROM robot|   \\
    Q   & $\mathit{ObstacleAhead}$   & \lstinline[language=SQL]|SELECT * FROM robot|\newline\phantom{W}\lstinline[language=SQL]|WHERE oAhead < min_forward_dist|   \\
    Q   & $\mathit{ObstacleLeft}$    & \lstinline[language=SQL]|SELECT * FROM robot|\newline\phantom{W}\lstinline[language=SQL]|WHERE oLeft < min_side_dist|  \\
    Q   & $\mathit{ObstacleRight}$   & \lstinline[language=SQL]|SELECT * FROM robot |\newline\phantom{W}\lstinline[language=SQL]|WHERE oRight < min_side_dist| \\
    U   & $\mathit{SetObstacles}$   & \lstinline[language=SQL]|UPDATE robot SET |\newline\phantom{W}\lstinline[language=SQL]|oAhead=:a, oLeft=:l, oRight=:r| \\\hline
\end{tabularx}    
    \caption{The ``query and command'' repository for the ROS case study.}
    \label{tab:case-studies:ros}
\end{table}

\subsubsection*{Behavioral Specification (CBTs)}

\begin{lstlisting}[
  float=t,
  label={lst:ros-bpjs},
  caption={The COBP version of the code in~\autoref{lst:ros-cpp}. The method \lstinline|bp.registerCBT| registers a CBT where the parameters are: the name of the CBT, the query name, and the behavioral specification of the CBT given a seed $r$.},
%   xleftmargin=.17\textwidth,
%   xrightmargin=.17\textwidth,
%   linewidth=0.7\textwidth
]
bp.registerCBT("Movement", "Robot", function(r) {
  while(true)
    bp.sync({ request:[
                move(0.3, 0),   /* move forward */
                move(0, -1.5),  /* turn left */
                move(0, 1.5)    /* turn right */] });
});

bp.registerCBT("Avoid obstacles: ahead", "ObstacleAhead", 
  function(r) {
    bp.sync({ block: move(0.3, 0), 
              waitFor: CTX.Ended("ObstacleAhead", r) });
});

bp.registerCBT("Avoid obstacles: left", "ObstacleLeft", 
  function(r) {
    bp.sync({ block: move(0, -1.5), 
              waitFor: CTX.Ended("ObstacleLeft", r) });
});

bp.registerCBT("Avoid obstacles: right", "ObstacleRight", 
  function(r) {
    bp.sync({ block: move(0, 1.5), 
              waitFor: CTX.Ended("ObstacleRight", r) });
});
\end{lstlisting}

\begin{lstlisting}[
  float=th!,
  label={lst:ros-extension},
  caption={Additional CBTs for the two new requirements: ``when the battery is low, the robot must reach the nearest power socket for recharging'', or ``if the robot has a package to deliver, it must first take it from the source location and then take it to its destination''.},
%   xleftmargin=.1\textwidth,
%   xrightmargin=.1\textwidth,
]
bp.registerCBT("GoToPowerSocket", "LowBattery", 
  function(r) {
    bp.sync({ request: bp.Event("NewTarget", 
                       { pos: socket_pos }) }); 
});

// The result of the $\mathit{Delivery}$ query has two properties of type $\mathit{Target}$ - source and target.
bp.registerCBT("Deliver", "Delivery", function(d) {
    bp.sync({ request: newTarget(d.source) }); 
    bp.sync({ waitFor: CTX.Ended("Target", d.source) });
    bp.sync({ request: newTarget(d.target) }); 
    bp.sync({ waitFor: CTX.Ended("Target", d.target) });
});

bp.registerCBT("GoToTarget", "Target", function(t) {
  while(true) {
    // calculate the moves for reaching the target.
    var path = calcPath(t.robot, t.pos);
    if(path.length == 0) {
      // The $\mathit{TargetReached}$ command ends the context by deleting t from table $\mathit{Target}$.
      bp.sync({ request: targetReached(t) }); 
      break;
    } else
      bp.sync({ block:   allMovesExcept(path[0]),
                waitFor: path[0]] });
  }
});
\end{lstlisting}

The code in~\autoref{lst:ros-bpjs} is the COBP implementation for the decision-making component. The ``Movement'' CBT specifies the possible moves of the robot, by constantly requesting to move forward, turn right, or turn left. The last three CBTs block linear and angular movements in case the movement will cause the robot to collide an obstacle. Each of these CBTs is bound to an obstacle query (i.e., $\mathit{ObstacleAhead},\mathit{ObstacleLeft},$ and $\mathit{ObstacleRight}$), \emph{blocking} the movement until the context ends. We note that the scan event is triggered by ROS and is not presented in this code. The execution engine of BPjs orchestrate the live copies, driving the robot around while avoiding the walls.\vspace{1em}

\subsection{A Short Discussion}
\label{sec:case-studies:ros:disscussion}
This case study demonstrated several advantages of the paradigm. 

\paragraph{Context activation vs. behavior}
The ``Avoid obstacles'' CBTs block the movement towards an obstacle as long as the context lasts. Another option to break the `blocking' is to wait for other movements, for example:
\begin{lstlisting}
bp.sync({ block:   move(0, 1.5), 
          waitFor: [move(0, -1.5),move(0.3, 0)] });
\end{lstlisting}
This implementation is a bad practice since it breaks the context abstraction. There is a difference between the sequence of events that triggered the activation/deactivation the context, and the behavior of a system during the context. Consider for example a new requirement that specifies that the robot can move backwards as well. In this case, the context will end, however the live copy that waits for other, specific movements --- will remain blocked, which may lead to a deadlock. 

\paragraph{Verification}
The $\mathit{block}$ idiom may lead to deadlocks in BP and COBP programs. For example, if our robot will reach a corner while facing it --- all of the possible moves will be blocked and our program will reach a deadlock. This bug, of course, is in the requirements, and can be solved by, e.g., allowing to move backwards. Yet, in more complex systems it might not be trivial to detect such deadlocks. Furthermore, the dynamic changes of the context may lead to additional unpredicted behaviors, a problem that is shared with the COP paradigm as well (elaborated in~\autoref{sec:related-work}). BPjs has a verification tool that we used to detect the deadlock. We elaborate on this tool in~\autoref{sec:case-studies:iot:discussion}. 

\paragraph{Higher level of incrementallity and agility}
The context idioms allow for adding new variations of the behavior under different contexts, thus improving the agility of the program. Consider for example a new requirement: ``when the battery is low, the robot must reach the nearest power socket for recharging'', or ``if the robot has a package to deliver, it must first take it from the source location and then take it to its destination''. Handling such requirements after the system is developed using a non-context-oriented approach, requires normally a redesign of the code. The COBP paradigm on the other hand, allows us to add these behavioral variation without changing the current CBTs, as demonstrated in~\autoref{lst:ros-extension}. The code assumes some changes to the schema --- adding a $\mathit{batteryLevel}$ attribute to the $\mathit{robot}$ table and two new tables, called $\mathit{Target}$ and $\mathit{Delivery}$. In addition, there are changes to the DAL --- adding queries and command for retrieving and manipulating the data of the new schema, as well as updating the effect function for triggering the new $\mathit{UPDATE}$ commands.

\begin{figure}
    \centering
    \fbox{\includegraphics[width=0.95\linewidth]{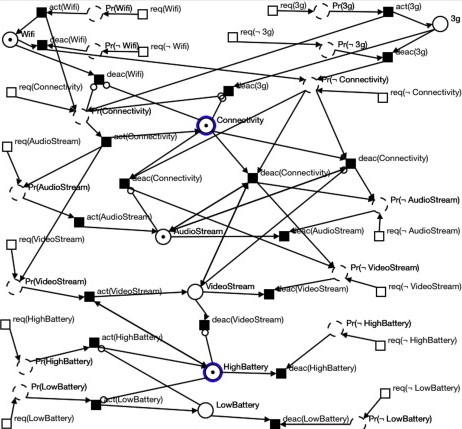}}
    \caption{A model of a specific context in a system and its relation to other contexts~\cite{Cardozo2015SemanticsConsistentActivation}. The user explicitly specifies the relations, such as: implications, requirements, conjunctions, disjunctions, causality, and more. In COBP, any order is possible, unless the requirements explicitly specify otherwise, which in that case, will be specified in the behavioral layer using CBTs.}
    \label{fig:copn}
\end{figure}

\paragraph{Comparison to COP}
As we explained in~\autoref{sec:intro}, the layers in COP are a language abstraction, grouping definitions of partial methods that implement some fragment of an aspect of the system behavior~\cite{Costanza2005LanguageConstructs}. The layers and the COBP queries share the key concept of grouping the behavioral variations. The name of the layer/query captures the context of these variations, and the definition of the partial methods are like the definitions of the CBTs. The differences though, reveal the some of the key differences between the two paradigms (that we discuss here and in~\autoref{sec:related-work}):

\begin{itemize}
\item \emph{Code organization:} One of the challenges in the design of context-aware systems is organizing code of crosscutting aspects~\cite{Salvaneschi2012ContextOrientedProgramming}. Context-dependent behaviors are often aspects that crosscut the application logic. While the partial methods in COP handle the aspects of the context-dependent behaviors, the other method parts (of these partial methods) may handle other aspects. Thus, it is difficult to organize the codebase without compromising the maintainability and the separation of concerns. In COBP, the alignment between the requirements and the specification, allows for a natural organization of the codebase. The queries are aligned to the contexts of the requirements (in our case study: robot, obstacles, etc.), and the CBTs are aligned to the requirements. 

\item \emph{Layer/Context activation:} In COP, the layer activation mechanism handles the composition of the partial methods of the activated layers. Such a composition is hard to handle since it is crucial to define (and when possible, to reason) the calling order of the partial methods as it may affect the behavior. To address this problem, several languages with formal semantics have been proposed for specifying the requirements of this composition. For example, Cardozo et al.~\cite{Cardozo2015SemanticsConsistentActivation} proposed a Petri net-based context model called CoPN (depicted in~\autoref{fig:copn}). Using this language, developer explicitly specify the relations between the different layers, where each relation can of a different type, such as: implications, requirements, conjunctions, disjunctions, causality, etc. These relations are usually not part of the requirements or the system specification, and are derived from the implementation of the system. Thus, it may be hard to change the model upon changes to the requirements or to the code. We further elaborate on COP semantics in~\autoref{sec:related-work}.

In COBP, all new live copies are simultaneously activated and advanced. If the requirements impose a specific order of events (as in the hot-cold example), then we will have a dedicated CBT for explicitly specify it. 
\end{itemize}

\section{Case Studies --- Smart Building}
\label{sec:case-studies:iot}
In this case study, we demonstrate how COBP can be used for developing reactive IoT systems, by developing the example of~\cite{Elyasaf2018LSC4IoT} using our BPjs-based implementation. We first list a set of requirements for a smart-building system and implement it. Next, we introduce additional requirements and refine the system specification, demonstrating the flexibility and agility offered by the paradigm. 

The initial requirements are as follows: 

\noindent \textit{Physical}: \textbf{R1}) The room types include offices, kitchens, and restrooms; \textbf{R2}) Each room has a motion detector and a smart light; \textbf{R3}) An Office has a smart air-conditioner; \textbf{R4}) Events are emitted when a motion starts or stops.

\noindent \textit{Behavioral}: \textbf{R5}) In all rooms, the light should be turned on once a motion is detected, and should be turned off if there is no motion detection for three minutes; \textbf{R6}) In office rooms, the air-conditioner should be turned on once a motion is detected, and should be turned off if there is no motion detection for three minutes; \textbf{R7}). In emergency, lights that are on must not be turned off.

We begin with identifying the context of the requirements, observing that while requirements R5 and R6 specify that the behavior depends on motion detection, the actual context of these requirements is that the room is empty/nonempty. Moreover, using a motion sensor for detecting occupancy is only one option that may change over time. Thus, we change these requirements slightly to depend on the occupancy of the rooms --- In $X$, the $Y$ should be turned on when there is someone in the room and turned off when the room is empty ($X$ is rooms/office rooms and $Y$ is light/air-conditioner, depending on the requirement). We also add the following requirement: \textbf{R8} A room is considered as empty if and only if it has no movement for more than three minutes.

\subsection{Implementation}
The possible events of the system are $\{\mathit{on}(r,d),\allowbreak \mathit{off}(r,d),\allowbreak \mathit{motionDetected}(r),\allowbreak \;\mathit{roomIsEmpty}(r), \mathit{roomIsNonempty}(r),\allowbreak \mathit{CTX.Ended}(q,c) \}$, where $\mathit{on}$ and $\mathit{off}$ are used to turn on/off a device $d$ in room $r$ (i.e., light or air-conditioner), $\mathit{motionDetected}$ pushed to the system by the motion sensor of room $r$,  $\mathit{roomIsEmpty}/\allowbreak\mathit{roomIsNonempty}$ are used to declare that room $r$ is empty/nonempty (respectively), and $\mathit{CTX.Ended}$ declares that $c$ is no longer the answer to query $q$ (i.e., the context $c$ has ended).

\subsubsection*{Context Specification}
\begin{figure}[th!]
    \centering
    \includegraphics[width=0.8\linewidth]{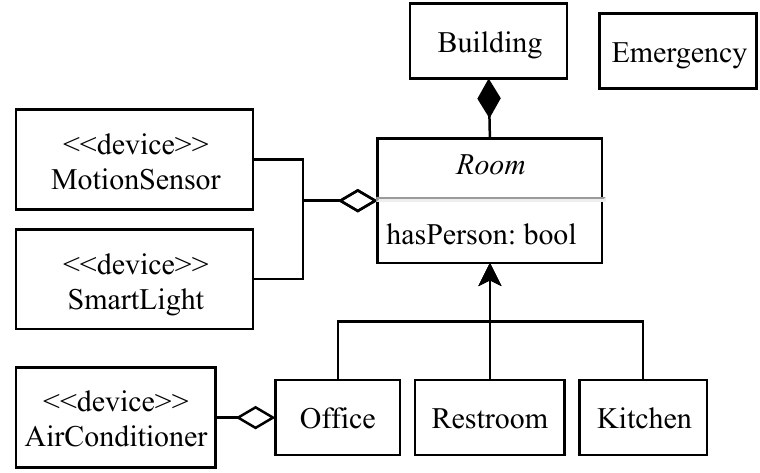}
    \caption{The context schema for the smart-building case study. Both physical and logical entities are represented in the schema.}
    \label{fig:methodology-cd-1}
\end{figure}

We begin with specifying the context in mind, in light of the given requirements, including physical entities and logical ones. \autoref{fig:methodology-cd-1} depicts the specification of the context schema (i.e., entities and the relationships among the them). It includes the following entities: Building, Room, Kitchen, Office, Restroom (originating from R1); the isEmpty attribute of Room (R5 and R6); and Emergency (R7). We further define the devices entities, i.e., MotionSensor, SmartLight (R2); and AirConditioner (R3). Of course this is just one of the ways to define the schema for these requirements. We adopt the UML class diagram notation for devising the schema, though other representations may be used as well. 

\begin{table}[th!]
    \centering
    \begin{tabularx}{1\linewidth}{@{}c@{\hspace{1ex}}l@{\hspace{1ex}}X@{}}
        \textbf{Type}   & \textbf{Name}   & \textbf{Command}   \\\hline
        Q   & $\mathit{Room}$   & \lstinline[language=SQL]|SELECT * FROM room|   \\
        Q   & $\mathit{Office}$   & \lstinline[language=SQL]|SELECT * FROM office|   \\
        Q   & $\mathit{Emergency}$    & \lstinline[language=SQL]|SELECT * FROM emergency|  \\
        Q   & $\mathit{EmptyRoom}$   & \lstinline[language=SQL]|SELECT * FROM room|\newline 
        \phantom{W}\lstinline[language=SQL]|WHERE isEmpty=1| \\
        Q   & $\mathit{NonemptyRoom}$   & \lstinline[language=SQL]|SELECT * FROM room|\newline 
        \phantom{W}\lstinline[language=SQL]|WHERE isEmpty=0| \\
        Q   & $\mathit{EmptyOffice}$   & \lstinline[language=SQL]|SELECT * FROM office|\newline 
        \phantom{W}\lstinline[language=SQL]|WHERE isEmpty=1| \\
        Q   & $\mathit{NonemptyOffice}$   & \lstinline[language=SQL]|SELECT * FROM office|\newline 
        \phantom{W}\lstinline[language=SQL]|WHERE isEmpty=0| \\
        Q   & $\mathit{NoMovement}$   & \lstinline[language=SQL]|SELECT * FROM room|\newline 
        \phantom{W}\lstinline[language=SQL]|WHERE timeFrom(lastMovement)|\newline 
        \phantom{WHERE}\lstinline[language=SQL]|> :seconds| \\
        U   & $\mathit{RoomIsNonempty}$   & \lstinline[language=SQL]|UPDATE room SET isEmpty=0|\newline
        \phantom{W}\lstinline[language=SQL]|WHERE id=:rId| \\
        U   & $\mathit{RoomIsEmpty}$   & \lstinline[language=SQL]|UPDATE room SET isEmpty=1|\newline
        \phantom{W}\lstinline[language=SQL]|WHERE id=:rId| \\
        U   & $\mathit{UpdateMovement}$   & \lstinline[language=SQL]|UPDATE room|\newline
        \phantom{W}\lstinline[language=SQL]|SET lastMovement=date('now')|\newline
        \phantom{W}\lstinline[language=SQL]|WHERE id=:rId| \\\hline
    \end{tabularx}
    \caption{The ``query and command'' repository for the smart building case study.}
    \label{tab:case-studies:iot1}
\end{table}

Once the schema is designed, we populate it with the initial predefined data. That is, adding buildings and rooms to the database and associating devices to specific rooms, and rooms to specific buildings. 

\subsubsection*{Data Access Layer}

$\mathit{QUERY}$ and $\mathit{UPDATE}$ are defined in~\autoref{tab:case-studies:iot1}. The effect of the $\mathit{motionDetected}$ event updates the timestamp of the last movement by calling $\mathit{UpdateMovement}$. The effect of $\mathit{roomIsEmpty}$ and $\mathit{roomIsNonempty}$ trigger a call to the corresponding update commands --- $\mathit{RoomIsEmpty}$ and $\mathit{RoomIsNonempty}$. 

\subsubsection*{Behavioral Specification (CBTs)}
The behavioral specification is given in~\autoref{lst:iot-bpjs}. The first two CBTs, $\mathit{Light\colon On/Off}$, are bound to queries $\mathit{Empty}\allowbreak \mathit{Room}$ and $\mathit{NonemptyRoom}$ (respectively), handling R5. The next two CBTs, $\mathit{Air\-conditioner\colon On/Off}$, are bound to the $\mathit{EmptyOffice}$ and $\mathit{NonemptyOffice}$ queries (respectively), handling R6. R7 is handled in the $\mathit{Emergency}\colon\allowbreak \mathit{Lights}$ CBT, and finally, the last two CBTs, $\mathit{Mark\,room\,as\,}\allowbreak \mathit{empty/nonempty}$ handle R8.

\begin{lstlisting}[
  float=t,
  label={lst:iot-bpjs},
  caption={The behavioral specification of the smart-building case study. The last two CBTs specify when to mark a room as empty/nonempty. The additional parameter in the last CBT provides the $\mathit{seconds}$ parameter to the $\mathit{NoMovement}$ query (see~\autoref{sec:case-studies:iot:discussion}).}
%   xleftmargin=.13\textwidth,
%   xrightmargin=.13\textwidth,
]
bp.registerCBT("Light: On", "NonemptyRoom", 
  function(room) {
    bp.sync({ request: on(room.light) });
});
bp.registerCBT("Light: Off", "EmptyRoom",
  function(room) {
    bp.sync({ request: off(room.light) });
});

bp.registerCBT("Air-conditioner: On", "NonemptyOffice", 
  function(office) {
    bp.sync({ request: on(office.airConditioner) });
});
bp.registerCBT("Air-conditioner: Off", "EmptyOffice", 
  function(office) {
    bp.sync({ request: off(office.airConditioner) });
});

bp.registerCBT("Emergency: Lights", "Room", 
  function(room) {
    while(true) {
      bp.registerCBT("Emergency: Light "+room.id,
        "Emergency", function(e) {
          bp.sync({ block: off(room.light),
                    waitFor: CTX.Ended("Emergency", e)}); 
          });
    }
});

bp.registerCBT("Mark room as nonempty", "EmptyRoom", 
  function(room) {
    bp.sync({ waitFor: motionDetected(room) });
    bp.sync({ request: roomIsNonempty(room) });
});
bp.registerCBT("Mark room as empty", "NoMovement", 
  {seconds: 3*60}, function(room) {
    bp.sync({ request: roomIsEmpty(room),
              waitFor: motionDetected(room) });
});
\end{lstlisting}

As development evolves, new requirements may arise. In our example these include:\\
\textit{Physical}: \textbf{R9}) Smart speakers are installed in all rooms (in addition to R2 devices); \textbf{R10}) Workers can be identified in the system (e.g., by a RF tag or a Bluetooth device), while visitors cannot.\\
\textit{Behavioral}: \textbf{R11}) If a worker enters a room, announce her name in the room's speaker; \textbf{R12}) During an emergency, turn on all lights.

\begin{figure}[t]
    \centering
    \includegraphics[width=0.9\linewidth]{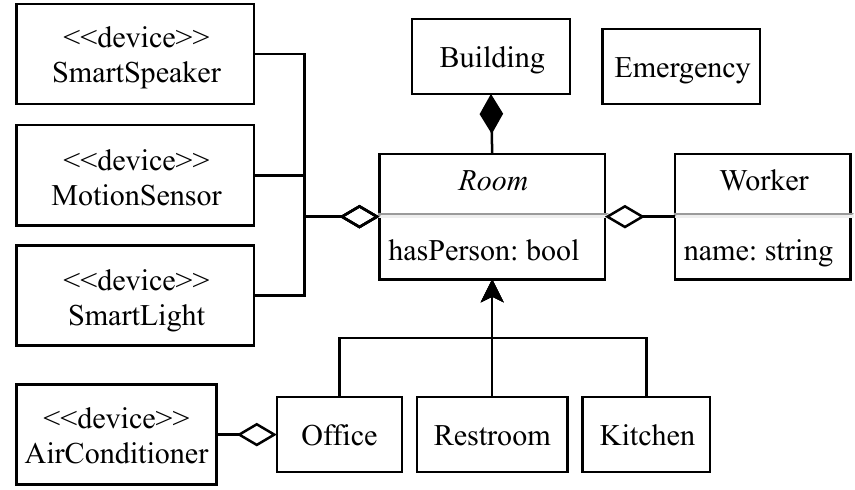}
     \caption{Schema for the new requirements}
    \label{fig:methodology-cd-2}
\end{figure}

To cope with these requirements, we can add object types that represent new aspects of the context. To the schema, we add a Smart Speaker data type (similarly to the other devices) (R9) and a Worker data type, both associated with Room (see~\autoref{fig:methodology-cd-2}). To the DAL we add commands for marking and un-marking that a worker has entered a room (R11) (similar to
the room occupancy update commands). We also modify the effect function for triggering them. For getting a view of all workers that are inside a room (R10), we add the $\mathit{WorkerInARoom}$ query, defined as \lstinline[language=SQL]|SELECT * FROM worker WHERE id IS NOT NULL|. Finally, we add the following CBTs: two for detecting entrance/leaving of workers (and triggering the appropriate marking/unmarking command using the effect function) (R10, R11); one for turning the lights on during an emergency (R12) (similar to the Emergency CBT in~\autoref{lst:iot-bpjs}; and one for announcing the worker name (demonstrated in~\autoref{lst:iot-bpjs-cont}) (R11).

Thanks to the incrementality feature of BP, no changes to the previous specification are needed, we only added elements to the schema, to the query and command repository, and to the behavioral specification. 

\begin{lstlisting}[
  float=t,
  label={lst:iot-bpjs-cont},
  caption={A new requirement --- announce workers names}
]
bp.registerCBT("AnnounceWorkerName", "WorkerInARoom", 
      function(worker) {
    bp.sync({ request: announce(worker.name) });
});
\end{lstlisting}

\subsection{A Short Discussion}
\label{sec:case-studies:iot:discussion}
\paragraph{Separation of concerns}
To determine the occupancy of a room, we created two CBTs that kept the $\mathit{isEmpty}$ attribute updated. It should be noted that there is simpler way to determine the occupancy --- remove the $\mathit{isEmpty}$ attribute and the update commands ($\mathit{RoomIsEmpty}$ and $\mathit{RoomIsNonempty}$) and change the queries $\mathit{EmptyRoom}$ and $\mathit{NonemptyRoom}$ to: 

\vspace{1ex}\noindent\begin{tabularx}{1\linewidth}{c l X}
    \textbf{Type}   & \textbf{Name}   & \textbf{Command}   \\\hline
    Q   & $\mathit{EmptyRoom}$   & \lstinline[language=SQL]|SELECT * FROM room WHERE|\newline  \lstinline[language=SQL]|timeFrom(lastMovement) > 3| \\
    Q   & $\mathit{NonemptyRoom}$   & \lstinline[language=SQL]|SELECT * FROM room WHERE|\newline \lstinline[breaklines=false,language=SQL]|timeFrom(lastMovement) <= 3|\\\hline
\end{tabularx}\vspace{1em}

While this solution is simpler and more concise, it breaks the separation of concerns and defines behavioral aspects in the data layer. Consider for example a change in the requirements, designed to improve the occupancy detection, where the occupancy is determined by two sensors and the thresholds for the sensors are different during working hours. By specifying the behaviors and the logical decisions in the CBTs only, we improve the agility of the system and achieve a better separation of concerns. 

\paragraph{Encapsulation}
One experience in developing COBP systems yielded an encapsulation rule of thumb. The CBTs that specify \emph{what} to do while a context is active, should be separated from the logic of \emph{when} to activate/deactivate a context, that ought to be encapsulated in dedicated CBTs. Consider for example future requirements for nonempty rooms, such as:  resume music, open the shutters, etc. The encapsulation of the \emph{when}-logic allows us to incrementally add simple \emph{what} CBTs, and it allows for future changes to the \emph{when} mechanism.

\begin{table}[b]
\setcounter{magicrownumbers}{0}
\centering
\begin{tabular}{c|@{\hskip 5pt}c@{\hskip 5pt}c@{\hskip 5pt}c@{\hskip 5pt}|@{\hskip 5pt}c@{\hskip 5pt}c@{\hskip 5pt}|}
 & \multicolumn{3}{@{\hskip 0pt}c@{\hskip 0pt}}{Setup} & \multicolumn{2}{@{\hskip 0pt}c@{\hskip 0pt}|}{Results} \\
 & \begin{tabular}[c]{@{}c@{}}\# of\\ rooms\end{tabular} & \begin{tabular}[c]{@{}c@{}}\# of\\ moves\end{tabular} & \begin{tabular}[c]{@{}c@{}}\# of\\ simulated\\ minutes\end{tabular} & \begin{tabular}[c]{@{}c@{}}\# of \\ states\end{tabular} & \begin{tabular}[c]{@{}c@{}}time \\ (seconds)\end{tabular} \\ \hline
\rownumber & 1 & 1 & 10 & 61 & 3.5 \\
\rownumber & 1 & 2 & 10 & 178 & 7.3 \\
\rownumber & 1 & 1 & 20 & 236 & 8.1 \\
\rownumber & 1 & 2 & 20 & 1496 & 42 \\
\rownumber & 2 & 1 & 10 & 869 & 29 \\
\rownumber & 2 & 2 & 10 & \multicolumn{2}{l|}{Out of memory} \\
\rownumber & 2 & 1 & 20 & 70001 & 177.4 
\end{tabular}
\caption{Experimenting with a non-context-aware verification algorithm on the smart-building example. The combination of two rooms and two movements resulted with ``out of memory'' exception, even though the sequence of events of each room is not influenced by the the events of other rooms. A context-aware algorithm could have verify the behavior of the only one room.}
\label{tab:verification-iot}
\end{table}

\paragraph{Verification}
As noted before, BPjs has a verification tool that we used for detecting undesired behaviors in this paper's examples. The tool traverses the state space of the program and validates certain aspects of the system behavior, such as: the absence of deadlocks or user-defined assertion. The input of this tool is the behavioral program and a set of assertions to validate. Once a violation is detected, the trace of events that led to the violation is returned to the user for inspection. 

The behavioral specification in our case depends on time and movements detection, both triggered by the environment. Since the environment must be simulated during the verification, we added one CBT for simulating movements in each room and one CBT for simulating time. We experimented with several configurations, varying in the number of rooms, the simulation length, and the number of movements. The experiments where conducted on a computer with an Intel i5-4570 CPU and 16GB RAM (using max. 12GB). The results, summarized in~\autoref{tab:verification-iot}, reveal that any combination of two rooms and two movements resulted with an ``out of memory'' exception, even though the sequence of events of each room is independent of the events of other rooms. The reason for this exception is that we did not modify the BPjs verification algorithm for context-related optimizations. Thus, it treats the context as a regular local variable of one of the b-threads (see our implementation in~\autoref{sec:cobp:implementation}). 
Yet, a simple context-aware optimization could have verify the behavior of only one room. More generally, while a b-program verification algorithm must consider all b-threads, in COBP it is sometimes enough to verify the CBT behavior instead of verifying the behavior of each live copy. We further discuss this in~\autoref{sec:related-work}.

\section{Related Work}
\label{sec:related-work}
Several modeling and programming languages include constructs that can be used for describing context. Harel's statecharts~\cite{Harel1987Statecharts} for example, introduced clustering for grouping behaviors under specific conditions. Below, we demonstrate how LSC constructs can also be used for modeling context. While COBP explicitly define context idioms, these languages do not. We identify three major advantages of using explicit context idioms. First, for achieving an alignment between the code/model and context-aware requirements, the language must include explicit ``first-class citizen'' idioms for referencing and changing the system context. Second, modeling context without explicit context idioms, may cause a tight coupling between the data specification and the behavioral specification, as we demonstrated in BP and as we demonstrate below in LSC. The context idioms proposed in this paper allow for a better separation-of-concerns by splitting the context-related design from the behavioral design, using the multilayered architectural pattern. Finally, the idioms and their formal semantics allow for improving verification algorithms, as we demonstrated in~\autoref{sec:case-studies:iot:discussion} and discuss below. 

\subsection{Context-Awareness in Scenario-Based Programming}
To the best of our knowledge, besides of the work of~\cite{Elyasaf2018LSC4IoT, Elyasaf2019UsingBehavioral} that we generalize here, there is no prior work on context for BP. Yet, there are are two, somewhat related works that we discuss here. 

The work on BP began with \emph{scenario-based programming} (SBP), a way to create executable specifications of reactive systems, introduced through the language of live sequence charts (LSC) and its Play-Engine implementation~\cite{Damm2001LSCsBreathing,Harel2003ComeLet}. As explained in~\autoref{sec:cobp:implementation}, while BP allows for behavioral specification only, the LSC language includes idioms for both behavioral and data specifications. The lifelines represent objects, possibly with data properties, allowing to share data among scenarios. A lifeline can either refer to a concrete object, or define a binding expression that is evaluated at runtime, meaning that a live-copy of the chart is instantiated whenever there is a new answer to the binding expression. Indeed here and in~\cite{Elyasaf2018LSC4IoT}, we utilize this trait for defining the context idioms as syntactic sugars of dynamic binding. Moreover, the Play-Engine specification splits each chart to a \emph{prechart} and a \emph{main chart}, where the prechart describes events (and conditions), that when they occur (or are satisfied), the system will attempt to execute the specification in the main chart. Thus, binding a chart to a context can be achieved by defining the context query inside the prechart area. While these two traits (dynamic binding and prechart) allow for sharing data among scenarios, as previously noted, explicit context idioms have several advantages. 

Atir et al.~\cite{Atir2008ObjectCompositionScenario} proposed to use the dynamic binding trait for extending LSC with idioms for specifying hierarchies between scenarios. Using hierarchies, they were able to abstract parts of the behaviors to different charts and activate them from several places. Atir et al. did not refer to context specification, however in some cases, context can be viewed (and implemented) as hierarchy. The COBP paradigm and the semantics we define here, extends this idea to other context types.


\subsection{Synthesis and Reasoning}
A recent literature review~\cite{Matalonga2017CharacterizingTestingMethods}, argued that the industry is currently concerned with the quality assurance of context-aware software systems (i.e., that the software will not fail upon context changes), while failing to cover aspects of functional assurance. The problem with covering all the possible variations of context during testing execution is not feasible. This is where formal methods and reasoning techniques may excel.

As we explained in~\autoref{sec:intro}, COP and COBP share the need for applying reasoning techniques on their model, in order to cope with unpredictable behavior that may occur due to the dynamic adaptation to context conditions. In fact, the BP paradigm faces this problem as well, due to the distributed nature of the paradigm, resulting in many tailored reasoning techniques that we now elaborate on.

Starting from the pioneering work of Harel, Kugler and Pnueli~\cite{Kugler2002SmartPlayout}, most of the synthesis and analysis tools for BP and SBP rely on the mathematically rigorous nature of the semantics in providing tools for running formal analysis and synthesis algorithms~\cite{Segall2007PlannedTraversablePlay,Kugler2005SynthesisRevisitedGenerating,Harel2010AcceleratingSmartPlay,BarSinai2018BPjs,Harel2011ModelcheckingBehavioral,Greenyer2017ScenarioTools,Harel2013ComposingProvingCorrectness,Piterman2006SynthesisReactive1Designs}. There are different approaches for verifying behavioral code: 

\begin{enumerate}
\item Harel et al.~\cite{Kugler2002SmartPlayout} translated the model to SMV~\cite{McMillan1993SMVSystem} and analyzed the equivalent SMV model. The method allows for symbolic model-checking that relies on a robust verification framework, though it requires a translation that may not be either sound or complete.

\item Bar-Sinai, Weiss and Shmuel~\cite{BarSinai2018BPjs} presented a verification tool for BPjs that we used for the examples of this paper (elaborated in~\autoref{sec:case-studies:iot:discussion}). The tool avoids the translation by directly traversing the state space of the program, as in NASA's JavaPathFinder (JPF)~\cite{Lindstrom2005ModelCheckingReal}. The method only allows for explicit model-checking. The advantage, however, is that it allows for model-checking any code because it uses the JavaScript interpreter in its state-space traversal. Bar-Sinai~\cite{BarSinai2020ExtendingBehavioralProgramming} compared the performance of this method to the performance of JPF (also summarized in~\autoref{tab:verification-comparison}). To compare, they verified with both tools different programs with various number of b-threads. JPF verified the program as a standard, non-BP program, therefore looking at all thread interleaving options, while BPjs only counted synchronization points as states. Not surprisingly, the JPF verification process was much longer, taking 85 seconds to verify a program containing three b-threads, visiting 438,568 states. While trying to verify a program with six b-threads, JPF ran out of memory after 8:31 minutes. BPjs was able to verify a program containing 10,000 b-threads in 10:67 minutes. 

\begin{table}[t]
    \centering
    \begin{tabular}{l|ccc}
        Tool    & B-Threads & States    & Time (msec) \\\hline
        JPF     & 3         & 438,568   & 511,000 \\
        JPF     & 6         & \multicolumn{2}{c}{out of memory} \\
        BPjs    & 25        & 27        & 2,205 \\
        BPjs    & 10,000    & 402       & 640,358 \\
    \end{tabular}
    \caption{A comparison between the performance of BPjs verification tool and NASA's JavaPathFinder (JPF). The formal semantics allow BPjs to count count as states only synchronization points, whereas JPF must look at all thread interleaving options.}
    \label{tab:verification-comparison}
\end{table}

\item \cite{Harel2013ComposingProvingCorrectness} used a hybrid approach that assumes that state explosion only comes from the b-threads composition, rather than their individual size. Given that, the space of each b-thread can be translated to a model, used by a model-checker for analyzing the composition of the b-threads. This method allows for a robust verification framework without manually creating a translator, though it only applies to systems with small b-threads (in terms of number of states). 

\item Maoz and Ringert~\cite{Maoz2015SynthesizingLegoForklift} used a tool for direct specification of models using SMV-like languages. They presented a case study of developing a software controller for a forklift robot using GR(1)~\cite{Piterman2006SynthesisReactive1Designs} synthesis tools. Their main observation is that extensions of the specification language with auxiliary variables and higher-level specification patterns support writing specifications with better confidence. On the other hand, with growing specification size, understanding reasons for synthesized behavior and for unrealizability turned out to be a major challenge. 
\end{enumerate}

In~\autoref{sec:case-studies:iot:discussion}, we demonstrated how the addition of context to the model improves its modularity, and how it can be used to improve the verification algorithm of~\cite{BarSinai2018BPjs}. Yet, such improvements, require an adaptation of each of the above approaches since all are designed under the assumption that the only protocol between b-threads is requests, blocking, and triggering of events. In practice, when people use BP with shared data they know that the model-checking technologies will not be applicable to their models. The formal COBP semantics presented here enable the change of this, by adding mechanisms to encode the state of the context data so that the model-checker is aware of the new form of inter-thread communication.

\subsection{Context-Oriented Programming}
BP and COBP are language-independent paradigms, with abstract semantics that are used and implemented by all BP languages. The evolution of COP was the other way around, where the first languages have been proposed before the semantics. Some COP implementations do not have formal semantics and require the translation of the actual code to some formal language, for applying reasoning techniques (see \autoref{sec:related:cop:abstract}). Semantics have also been proposed for specific COP languages (see \autoref{sec:related:cop:specific}). Finally, some COP implementations allow for developing the system in a modular manner (see \autoref{sec:related:cop:modular}), where part of the system (for example, the layer-activation constraints) is formally specified, and the rest is defined using the underlying programming-language constructs (e.g., Java, Python, etc.). 

All the different COP languages and semantics elaborate below do not provide a formal semantics for the complete program (i.e., the business logic and its relation to the context). Thus, reasoning techniques can be applied only to the formally-specified parts, as opposed to COBP, where the model can be both executed and reasoned. 

\subsubsection{Language-Independent Semantics}
\label{sec:related:cop:abstract}
In \autoref{sec:case-studies:ros:disscussion} we elaborated on CoPN, a Petri net-based language for modeling run-time context activations~\cite{Cardozo2015SemanticsConsistentActivation}, allowing for both reasoning the model and using it at run-time. \cite{Schippers2010GraphbasedOperational} proposed another approach for representing program states as graph nodes, connected by their relations. This approach allows for simulating context-oriented programs, though it is not possible to reason the consistency of the activation nor use it for run-time context compositions. 

\subsubsection{Language-Specific Semantics}
\label{sec:related:cop:specific}
Costanza and D'Hondt~\cite{Costanza2008FeatureDescriptionsContext} extended the ContextL language with declarative constraints on layers. The extension simplifies the enforcement of the constraints and allows the programmer to interactively fulfill unmet constraints. 
The authors of EventECJ proposed to manually specify transition-rules between different contexts with finite-state automata~\cite{Kamina2011EventCJContextoriented, Aotani2011FeatherweightEventcjCore}. Another solution, presented for Subjective-C, use a DSL for expressing layer dependencies~\cite{Gonzalez2008ProgrammingAmbience}. While these semantics are language-specific (and there more), some of them allow for reasoning both the context activation and the behavioral variation.

\subsubsection{Modular Languages}
\label{sec:related:cop:modular}
A modular approach has been proposed in~\cite{Tomoyuki2016TowardsModularReasoning}, where layer interfaces define a contract between the layers and the classes. Each layer must implement one or more layer interfaces, and each class allows layer interfaces. Implementation means that the behavior of the layer satisfies the layer interfaces it implements, and allowing an interface layer means that the class satisfies the specifications for the allowed layer interface. This change allows for applying a modular reasoning in the layer activation mechanism for composing the partial methods in the activated layer.




\section{Concluding Remarks}
\label{sec:concluding}
This paper introduced \emph{Context-Oriented Behavioral Programming} (COBP) --- a novel paradigm for developing context-aware systems --- built on top of the behavioral-programming paradigm. The paper aimed at formally defining the paradigm and introducing it through different examples and discussions. As described in~\autoref{sec:related-work}, more research is required on adapting current reasoning techniques for effectively applying them on COBP specifications. Also, further user studies and case studies are needed for comparing the paradigm to other paradigms and for evaluating additional aspects of the paradigm. Specifically: 

\begin{itemize}
\item \emph{Comparing COBP and BP.} The advantages (and possible disadvantages) of COBP over BP can be evaluated by many metrics, such as improved specifications in terms of readability, understandability, maintainability, incrementality, succinctness. Moreover, COBP provides more design patterns than BP. For example, in BP, conflicting behaviors can be arbitrated by specifying priorities, whereas in COBP they can also be defined in different contexts.

\item \emph{Comparing COBP and COP.} Both COBP and COP allow for context-aware programming, though they are fundamentally different from each other. Some of the issues that were presented here from the perspective of COBP, have been addressed and discussed by others from the perspective of COP. An in-depth comparison between the two paradigms is needed for comprehending the differences between them.

\item \emph{Evaluating methodology aspects.} We present several approaches for specifying COBP programs. The different examples demonstrated good and bad practices, discussing their advantages and disadvantages. Yet, these practices impact both qualitative and quantitative aspects of the specifications, raising the need for a user study that will evaluate these aspects.
\end{itemize}

\section*{Acknowledgments}
The author would like to thank Gera Weiss for his invaluable help with this paper.

\bibliographystyle{plain}  
\bibliography{gdrive-library}  
\end{document}